\documentclass[fleqn, usenatbib, usedcolumn]{mnras}
\usepackage[dvipsnames]{xcolor}
\usepackage{newtxtext,newtxmath}
\usepackage{natbib}
\usepackage[T1]{fontenc}
\usepackage{xcolor}

\usepackage{graphicx}	% Including figure files
\usepackage{subcaption}
\usepackage{soul}
\usepackage{amsmath}	% Advanced maths commands
\usepackage{fontawesome}
\usepackage{hyperref}
\hypersetup{
    colorlinks,
    linkcolor={red!50!black},
    citecolor={blue!50!black},
    urlcolor={blue!80!black}
}
    
\DeclareRobustCommand{\VAN}[3]{#2}
\let\VANthebibliography\thebibliography
\def\thebibliography{\DeclareRobustCommand{\VAN}[3]{##3}\VANthebibliography}

\definecolor{scc}{rgb}{0.13, 0.67,  0.8}
\definecolor{scc2}{rgb}{0.63, 0.67,  0.8}
\title[3D Detection and Characterisation of ALMA Sources through Deep Learning]{3D Detection and Characterisation of ALMA Sources through Deep Learning}

\author[M. Delli Veneri et al.]{
Michele Delli Veneri,$^{1, 2}$\thanks{E-mail: delliven@na.infn.it, micheledelliveneri@gmail.com}
{\L}ukasz Tychoniec,$^{3}$\thanks{E-mail: Lukasz.Tychoniec@eso.org }
Fabrizia Guglielmetti $^{3}$,
Giuseppe Longo $^{2}$,
Eric Villard $^{3}$ 
\\
$^{1}$INFN Section of Naples, Napoli, via Cintia, 1, Italy, 80126\\
$^{2}$Department of Electrical Engineering and Information Technology, 
	University of Naples "Federico II", Via Claudio, 21, 80125 Naples NA, Italy\\
$^{3}$ESO, Karl-Schwarzschild-Straße 2, 85748 Garching bei München\\
$^{4}$Department of Physics "Ettore Pancini", University of Naples "Federico II"
	, Via Cintia, 1, Italy, 80126\\
}

\date{Accepted 2022 November 9. Received 2022 November 9; in original form 2022 August 12}

\pubyear{2022}

\begin{document}
\label{firstpage}
\pagerange{\pageref{firstpage}--\pageref{lastpage}}
\maketitle

\begin{abstract}
	We present a Deep-Learning (DL) pipeline developed for the detection and characterization of astronomical sources within simulated Atacama Large Millimeter/submillimeter Array (ALMA) data cubes. The pipeline is composed of six DL models: a Convolutional Autoencoder for source detection within the spatial domain of the integrated data cubes, a Recurrent Neural Network (RNN) for denoising and peak detection within the frequency domain, and four Residual Neural Networks (ResNets) for source characterization. The combination of spatial and frequency information improves completeness while decreasing spurious signal detection. To train and test the pipeline, we developed a  simulation algorithm able to generate realistic ALMA observations, i.e. both sky model and dirty cubes. The algorithm simulates always a central source surrounded by fainter ones scattered within the cube. Some sources were spatially superimposed in order to test the pipeline deblending capabilities. The detection performances of the pipeline were compared to those of other methods and significant improvements in performances were achieved. Source morphologies are detected with subpixel accuracies obtaining mean residual errors of $10^{-3}$ pixel ($0.1$ mas) and  $10^{-1}$ mJy/beam on positions and flux estimations, respectively. Projection angles and flux densities are also recovered within $10\%$ of the true values for $80\%$ and $73\%$ of all sources in the test set, respectively. While our pipeline is fine-tuned for ALMA data, the technique is applicable to other interferometric observatories, as SKA, LOFAR, VLBI, and VLTI.
\end{abstract}


%%%%%%%%%%%%%%%%%%%%%%%%%%%%%%%%%%%%%%%%%%%%%%%%%%

%%%%%%%%%%%%%%%%% BODY OF PAPER %%%%%%%%%%%%%%%%%%

\section{Introduction}\label{sec:introduction}
In the last two decades, astronomical datasets underwent a rapid growth in size and complexity thus pushing Astronomy in the big data regime \citep{longo, baron, pesenson, Brescia2021}. 
Traditional approaches such as interactive data reduction and analysis are laborious due to the size and the complexity of the data hence the ability of machine learning methodologies, both supervised and unsupervised to cope with very complex data, has been extensively exploited by the community to solve a wide variety of problems spanning all aspects of the astronomical data life, from instrument monitoring to data acquisition and ingestion, to data analysis and interpretation \citep{becker_2020, Kovaevi2020, Huertas_Company_2018, MargalefBentabol2020, Lanusse2021, Morningstar2019, swere, Zhao, Lin2022, Duarte2022, Cheng2020}.
This has also led to the implementations of many deep learning-based pipelines in the field of Astrophysics. To quote only some among the most recent and interesting applications,
\cite{Yi2022} utilized a custom deep learning model to detect Low Brightness Galaxies within SDSS images. Their architecture is composed of three models: a ResNet50 to extract features from the images, a CNN to regress the class probabilities for the galaxies within the images, and another CNN to predict the coordinates of the galaxies within the images. As loss functions, they employed the Binary Cross Entropy loss for the classification problem and a probability-weighted mean squared loss for the regression problem. 
\cite{Akhazhanov2022} proposed a novel deep learning-based pipeline for detecting quadruply lensed quasars. The pipeline was trained on simulated observations comprising both lensed and non-lensed quasars and is composed of several deep learning models combined together to solve a binary classification problem: the lower-dimensional representation extracted from four Variational Autoencoders, two ResNets, one with rectangular convolution, one with polar, and a U-Net-like model. 
\cite{Goode2022} presented a novel pipeline called Removal of BOgus Transients (ROBOTs) in order to assess transient candidates within the Deeper, Wider, Faster (DWF) program. One of the main bottlenecks of DWF, is the time required to assess candidates for rapid follow-up and manual inspection prior to triggering space-based or large ground-based telescopes, the authors create a pipeline to address this main bottleneck. The pipeline employs a combination of a CNN and a CART: the CNN is used to classify the input images, while the CART is used to further classify  candidate light curves identified by the CNN. In fact, the CNN algorithm is used to determine the quality of individual data points rather than that of the light curves, and thus the CART is used to incorporate the temporal information within the decision framework. As we will demonstrate, their strategy of utilizing extra information in the temporal domain to further classify candidates is similar to how we use frequency information to deblend the potential candidates found by the AutoEncoder in our pipeline (see Sec. \ref{subsec:pipeline} step 8). 
\cite{pearson2019} utilized a CNN trained with both real observations from the SDSS \citep{sdss} and simulated observation from EAGLE \citep{eagle} to identify galaxy mergers. By comparing the performances on both real and simulated data, they concluded that a CNN trained on simulated data could be used to detect mergers in real observations.
Also in the field of radioastronomy, some interesting  examples can be found. 
For instance, \cite{Bowles2020} introduced attention-gated networks for radio galaxy classification demonstrating that these networks can perform similarly to CNN-based classifications while utilizing significantly fewer trainable parameters and thus reducing the complexity of the model and the risk of overfitting. Moreover, they employed the produced attention maps \citep{attention_maps}  to help the interpretation of the results.
\cite{Zeng2020} proposed a novel deep-learning pipeline, Concat COnvolutional Neural 
Network (CCNN), for selecting pulsar candidates within the Commensal Radio Astronomy FasT Survey \citep{Li2018}. The main idea behind their pipeline is to use several specialized CNNs to extract the low-dimensional latent information from the pulse profile, the Dark Matter profile, the frequency versus phase plot, and the time versus phase plot. The latent spaces are then concatenated and fed to a Multi-Layer Perceptron that performs a binary classification task. 
\cite{VafaeiSadr2019} employed a CNN to detect point sources within images. The CNN is trained on simulated maps with known point source positions and learns how to amplify the Signal to Noise Ratio (SNR) in the input map. 
The map is then converted into a catalogue using a dynamic blob detection algorithm. The authors compared their pipeline performances with PyBDSF (a classical, widely used peak detection algorithm, \cite{PyBDSF}), obtaining better results on all metrics.
\cite{Lukic2019} presented a novel deep learning model called ConvoSource, which utilizes 2D Convolutions to map the input observations to the target signal map using a binary cross-entropy loss function. The model performance was found to be similar to that of PyBDSF. Both \cite{VafaeiSadr2019} and \cite{Lukic2019} use as input images already correlated data and thus rely on the \textsc{CLEAN} algorithm \citep{hogbom}: an assumption which as we shall discuss below, may affect the final performances.
\cite{Connor2022} developed a novel deep learning pipeline called POLISH for image super-resolution and deconvolution of radio interferometric images. Their model aims to learn the mapping between the input "dirty" images and the target sky model images. Their architecture is a modified version of the Wide Activation for Efficient and Accurate Image Super-Resolution \citep{Yu2018} capable to allow for high-dynamic range, multifrequency output and uncertainty in the PSF. 
To test the performances of their model they generated DSA-2000 \citep{dsa2000} simulated data, and compared the performances of their model on the dirty images with \textit{source extractor} applied on cleaned images. The cleaning was performed with the \textsc{CLEAN} algorithm \citep{hogbom}. POLISH  achieved both better mean peak signal-to-noise ratio (PSNR) and better structural similarity index (SSIM) as well as a lower fraction of false detections when compared to the true sky models. 
\cite{Schmidt2022} presented a deep-learning architecture for reconstructing incomplete Fourier data in radio interferometric images. By reconstructing missing data in the visibility space, a clean sky model image can be obtained by simply taking the inverse 2D Fourier transform of the reconstructed data. In order to train and test the performances of their architecture, the authors generated simulated VLBI $u-v$ maps both noiseless and noisy, corresponding for both point-like and Gaussian sources and compared the results of their model with those by \textit{wsclean} \citep{offringa-wsclean-2014}. 

The direct reconstruction of the sky model from the $u-v$ map has been tackled with classical statistical approaches by several authors \citep{Honma2014, Kuramochi_2018, Akiyama_2017} but, to the best of our knowledge, \cite{Schmidt2022} were the first to tackle the problem using a Deep Learning approach. 

Finally, \cite{Rezaei2021} presented a novel deep learning pipeline (DECORAS) for detecting and characterizing sources in VLBI simulated images. Their pipeline consists of two Deep Convolutional Autoencoders (similar in scope and structure to the first deep model of our pipeline, i.e. Blobs Finder) used to detect sources within the images and an XGboost model for the regression of source peak surface brightness. Other morphological parameters are derived by fitting 2D Gaussian models to the sources. The authors train and test their pipeline on simulated VLBI simulations and directly compare with \textsc{blobcat} \citep{blobcat}, a source extraction software that utilizes the \textit{flood fill} \citep{scikit-learn} algorithm to detect and catalog blobs, showing that the two methods have similar completeness levels for $SNRs <= 5.5$. While for $SNRs > 5.5$,  DECORAS outperforms \textsc{blobcat} by a factor of two.
\medskip

Last but not least we must mention \textsc{Sofia} \citep{Serra} which -even though not based on DL  is a state of the art flexible line finding algorithm capable of detecting and parametrizing HI sources within 3D radio data cubes. When employing the smooth and clip algorithm \textsc{S + C} , meaningful emission in the cube is detected by smoothing the data cube with 3D kernels specified by the user at multiple angular and velocity resolutions. At each resolution, voxels are detected if their absolute value is above a threshold given by the user (in noise units). The final mask is the union of the masks constructed at the various resolutions. The algorithm was updated and renamed \textsc{Sofia-2} \citep{sofia}, rewritten in the C programming language while making use of OpenMP for multi-threading of the most time-critical algorithms. \textsc{Sofia-2} is substantially faster and comes with a much reduced memory footprint compared to its predecessor.
\medskip

Our paper is arranged as follows:
in Section~\ref{sec:dl_and_alma} we introduce the image reconstruction problem in radio astronomy and how it is solved for ALMA images.
In Section~\ref{sec:simulations} we describe the simulation algorithm that we used to generate realistic ALMA observations. The simulation algorithm is used to train and test our pipeline. The simulation parameters and the characteristics of the output cubes and the sources within them are explained.
In Section~\ref{sec:dl} the architectures of the employed deep learning models of our pipeline, together with 
a complete data flow are described. The data flow is used to  explain the inner workings of the pipeline, as well as a description of the training strategies adopted for all the models. 
In Section~\ref{sec:source_detection} we discuss the pipeline performances in detecting and characterizing sources within the test set. We compare the detection performances with those of two traditional methods \textsc{blobcat} \citep{blobcat}, \textsc{Sofia-2} \citep{sofia}, and another deep learning pipeline \textsc{decoras} \citep{Rezaei2021}. 
Finally, in Section~\ref{sec:conclusions} we discuss our results, draw our conclusions, and lay the prospect for future work. 
Both the simulations code (ALMASim \href{https://github.com/MicheleDelliVeneri/ALMASim}{\faGithub}) and the pipeline (DeepFocus \href{https://github.com/MicheleDelliVeneri/DeepFocus}\faGithub) are written in Python \citep{python}, and all software has been made publicly available through GitHub in order to allow for further developments, testing and reproducibility.

\section{Deep Learning and ALMA}\label{sec:dl_and_alma}

Interferometric radio observatories such as ALMA, after the initial correlation and calibration of the raw signals detected by the antennas in Fourier space, provide data in the form of measurement sets, which contain information of visibilities for each baseline. In imaging, cubes are retrieved from the measurement sets through an inverse Fourier transform. 
Data cubes are composed by a series of images (chracterized by their coordinates on the celestial sphere) each taken at a different frequency. Frequency channels are usually contiguous and their separation is determined by the frequency resolution (or sensitivity) of the observatory. 
The extraction of the astrophysical signals from the data cubes requires the solution of an ill-posed inverse problem: 
\begin{equation}
	I^{D}(x,y) = R \times I(x,y) + n
\label{n1}
\end{equation}
where $I^{D}(x,y)$ are the observed measurements, $I(x,y)$ are the unknown signals, $R$ is a degradation operator due to the response function of the observatories and $n$ is the additional noise propagated from the observation process to the calibrated signal.
$R$, also known as \textit{forward operator}, takes up the very complex underlying physical phenomena involved in the observational process. The deconvolution process ideally provides the noiseless observation $I(x,y)$ from the observed measurements and it is traditionally performed making assumptions about the structure of both the signal and of the forward operator.
Many attempts have been made at solving this problem using Machine Learning (ML) based approaches \citep{Bowles2020, Zeng2020, Schmidt2022, Rezaei2021, Connor2022}. In this paper, we present a deep-learning-based pipeline for the detection and characterization of sources within ALMA ``uncleaned`` or ``dirty`` calibrated data cubes.
In a first order approximation, a dirty cube represents the inverse Fourier transform of the observed \textit{visibilities} convolved with the instrumental point spread function (dirty beam). Where for Visibilities we mean the recorded complex values of the interference pattern provided by each antenna pair \citep{thompson2008interferometry}. Since for every observation, the number of sampled visibilities is forcefully limited, a direct inversion of Eq.~\ref{n1} is not feasible for reconstructing the sky brightness.
The most popular deconvolution and cleaning technique for image reconstruction in the radio domain is \textsc{CLEAN} \citep{hogbom}. In CLEAN, given the point-spread function, point sources are identified and subtracted from the dirty image using an iterative process. The identified point sources are recorded in the model image. When all point sources are removed from the dirty image, the remaining dirty image should consist only of noise. A multiscale \textsc{CLEAN} approach extends the work of \cite{hogbom} allowing point sources to be Gaussian distributed instead of delta functions \citep{MCLEAN}.
ALMA has been a game changer in high resolution aperture synthesis imaging \citep{alma_2015}, and a development roadmap \citep{Carpenter_2022} has been approved with the goal of meeting its high standards. One of its main objectives is to broaden the instantaneous bandwidth of the receivers, upgrading the correlator to process the entire bandwidth. As a consequence, ALMA sensitivity and observing efficiency will improve, producing imaging products at least two orders of magnitude larger than the current cube size (which already resides in the GB regime). As a result of these upgrades, ALMA cube imaging will become a very demanding task.

In this work we introduce a new approach to the image reconstruction problem capable to overcome some of the limitations introduced by the \textsc{CLEAN} algorithm. More in details,  \textsc{CLEAN}'s iterative cleaning procedure optimizes the best possible image reconstruction employing a minor cycle (or deconvolver) operating in the image domain and a major cycle (or imager) to handle residuals from the observed data and the estimated model image (through a transformation from data and image domains). Each cube  undergoes a time-consuming cleaning procedure which is demanding for current and future radio interferometers \citep{Carpenter_2022}. 
Moreover, by working on each slice independently, \textsc{CLEAN} does not allow for correlating information between pixels along the frequency axis of the cube, thus leading to the creation of artefacts in the cleaned cube. For example, a noise peak would be deconvolved several times with the instrumental PSF and the recovered delta function would be convolved with the \textit{clean beam} (i.e. a Gaussian approximation of the PSF) thus producing a structure morphologically similar to the actual sources that underwent the same iterative deconvolution process. 

We propose, instead, a deep-learning-based pipeline developed with the goal to detect emission lines in ALMA cubes while speeding up traditional methods and reducing spurious signal detection. For the aforementioned reasons, our pipeline uses as input information, 'dirty' calibrated ALMA cubes that have not undergone any prior deconvolution.
The main novelty of our proposed pipeline with respect to the previously cited architectures 
\cite{VafaeiSadr2019,Lukic2019, Connor2022, Schmidt2022, Rezaei2021} is that we combine spatial and spectral (frequency) information to detect and characterize sources. Utilising frequency information can both help in deblending spatially blended sources in the integrated images (obtained integrating the cubes along the frequency axis), and help in the recovery of faint sources.
Our deep learning pipeline can be decomposed into six logical steps:
\begin{enumerate}
	\item 2D source detection is performed on the integrated cubes (along frequency) using a Deep Convolutional Autoencoder (hereafter Blobs Finder);
	\item the detections (blobs) found by the autoencoder are used to extract spectra which are then fed to a Recurrent Neural Network, capable to perform spectral denoising (Deep GRU). All the peaks in the spectra are fitted with 1D Gaussian distribution;
	\item sources are spectrally focused by cropping a $64 \times 64$ pixels box around their centres (found by Blobs Finder) in the image plane and by integrating within their frequency emission ranges found in the previous step. False positives are detected in this step;
	\item focused images are then fed to three dedicated ResNets that regress the sources morphological parameters: full-width half maxima in the $x$ and $y$ directions and the projection angle $\theta$;
	\item the found morphological parameters are used to construct a 3D models of the sources, which are used to mask the cubes and produce line emission images and mean continuum images. Once the continuum is subtracted from the line emission images, the latter are finally fed to a specialized ResNet that regresses the source flux densities.
\end{enumerate}	

\section{Simulations}\label{sec:simulations}
To train and test the capabilities of the proposed pipeline, we needed thousands of ALMA model and dirty cube pairs. 
We generate our own simple but realistic simulations of ALMA observations by combining python and bash scripting with the Common Astronomy Software Application (CASA) v. 6.5.0.15 \citep{casa} python libraries. 
The need to use simulated data instead of real ALMA observations arises from several practical necessities: a) in order to evaluate the pipeline's performance and reliability and its dependence from sources observational and morphological parameters, we need full control over the data properties; b) in order to assess the reconstruction quality of our pipeline and its ability to solve the deconvolution problem without relying on \textsc{CLEAN} deconvolution, we need noiseless sky observations which are unattainable for real observations. The closest alternative to the use of simulated cubes would be to use \textsc{CLEAN} to produce deconvolved cubes of ALMA observations in which source detection had been performed. These deconvolved cubes, however, would be dependent on CLEAN solution (of the deconvolution problem), thus biasing our models.
Given that the main foreseen scientific application of the pipeline presented in this work is the search of serendipitous sources within high redshift ($z > 3$) ALMA observations, we simulate, in each cube, a central primary source with a given SNR (simulating this way a realistic target observation, in the phase centre of the antennas array),
surrounded by less bright serendipitous sources that can occupy any position in the cube.
To generate 3D sources, we combine 2D Gaussian Components in the spatial plane with 1D Gaussian components (emission lines) in frequency space. For each source, first, the emission line is generated by sampling from a uniform distribution of line parameters, i.e., the line position or central
frequency, the line amplitude or the value in dimensionless units of the peak, and the line FWHM in number of frequency channels.
Once the parameters have been sampled, the line is generated through the following equation:
\begin{equation}
	l(z) = a e^{- \frac{(z - z_{cen})^2} {2 (FWHM_z / 2.35482)^2}}
\end{equation}
where $a$ is the line amplitude, $z_{cen}$ is the central frequency and $FWHM_z$ is the line full width half maximum.

Then a 2D Gaussian profile is generated by sampling from a uniform distribution of source parameters, i.e., the source position $x$, $y$, the source FWHMs in both the $x$ and $y$ planes $FWHM_x$ and $FWHM_y$, the source 
projection angle $\theta$, the source amplitude $a$ and the source spectral index $\sigma$, i.e. the dependence of  the source radiative flux density on frequency. 
Once the parameters have been sampled, the 3D Gaussian profile is generated
through the following set of equations:
\begin{equation}
\begin{array}{l}
	g(x, y, z) = 10^{(log_{10}(a) + spid \times log_{10}(v_1 / v_2))} \cdot\\
	\quad \cdot e^{- \frac{((x_c - x) / FWHM_x)^2 + ((y_c - y) / FWHM_y)^2}{2}}  \quad\\
\end{array}
\end{equation}
where
\begin{equation}
\begin{array}{l}
	x_c = x \cdot cos(pa) - y \cdot sin(pa)\\
	y_c = y \cdot sin(pa) + y \cdot cos(pa)\\
	v_1 = 230 \cdot 10^9 - (64 \cdot 10^6)\\
	v_2(z) = v_1 \cdot z \cdot 10^6\\
	G(x, y, z) = g(x, y, z) + l(z) * g(x, y, z)
\end{array}
\end{equation}
Each model cube is thus created by first generating a central source and then a random number between $2$ and 
$5$ additional sources such that their emission peaks are lower or equal to that of the central source. The full list of parameters, their units, and the ranges from which they are sampled, are shown in Tab~\ref{tab:Gaussian_parameters}.

\begin{table}
	\centering
	\caption{Sampling intervals of the model source parameters. Sources are generated by randomly sampling
	from the outlined uniform distributions. The first column shows the parameter name, the second the range
	from which the parameter values are sampled, and the third the units.}
	\label{tab:Gaussian_parameters}
	\begin{tabular}{lcc} % four columns, alignment for each
		\hline
		 Parameter Name                 & Range        & Unit       \\
		\hline
		Number of components            & [2 -- 5]     & --            \\
		Amplitude of 2D Gaussian compoment & [1 -- 5]     & arbitrary     \\
		FWHM of the 2D Gaussian component  & [2 -- 8]     & pixel         \\
		Spectral index                  & [-2 -- 2]    & --            \\
		Position in the xy plane          & [100 -- 250] & pixel         \\
		Position angle                  & [0 -- 90]    & deg           \\
		Line amplitude                  & [1 -- 5]     & arbitrary     \\
		Line center                     & [10 -- 110]  & chan          \\
		Line FWHM                       & [3 -- 10]    & chan          \\

		\hline
	\end{tabular}
\end{table}
All source parameters are stored in a .csv file in order to use them as targets for the ResNets in the 
last stages of the Deep Learning source detection pipeline (see Sec.~\ref{subsec:pipeline}).

We  feed the sky models to the \textit{simobserve} task to simulate interferometric measurements sets through a
series of observing parameters. The full list of parameters is outlined in Tab~\ref{tab:simobserve_parameters}.
\begin{table}
	\centering
	\caption{Full list of \textit{simobserve} parameters to generate the measurement sets, i.e.~the interferometric
	visibility data cubes. The first column shows
	the parameter name, while the second our chosen value. The first and second columns show the parameter name and our selected value, respectively.}
	\label{tab:simobserve_parameters}
	\begin{tabular}{lc} % four columns, alignment for each
		\hline
		 Parameter Name   & Value\\
		\hline
		inbright          & $0.001$ Jy / Pix \\
		indirection       & J2000 03h59m59.96s -34d59m59.50s\\
		incell            & $0.1$ arcsec\\
		incenter          & $230$ GHz\\
		inwidth           & $10$ MHz\\
		integration       & $10$ seconds\\
		totaltime         & $2400$ seconds\\
		thermalnoise      & tsys-atm\\
		user\_pwd          & $0.8$\\
		\hline
	\end{tabular}
\end{table}

\begin{figure*}
	\centering
	
	\begin{subfigure}[b]{1\columnwidth}
		\includegraphics[width=\textwidth]{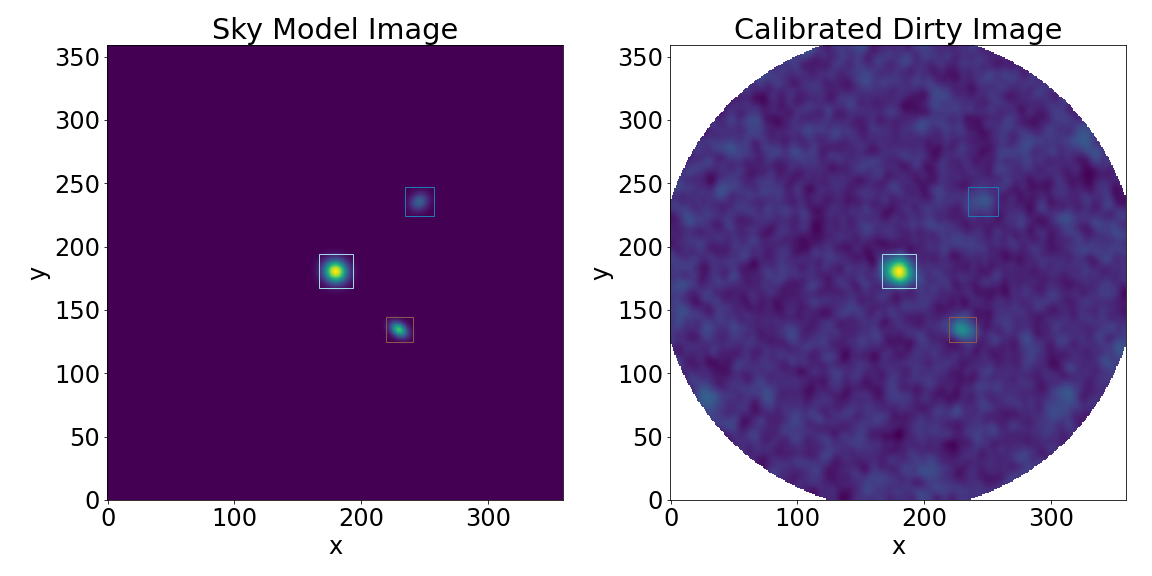}
	\end{subfigure}
	\hfill
	\begin{subfigure}[b]{1\columnwidth}
		\includegraphics[width=\textwidth]{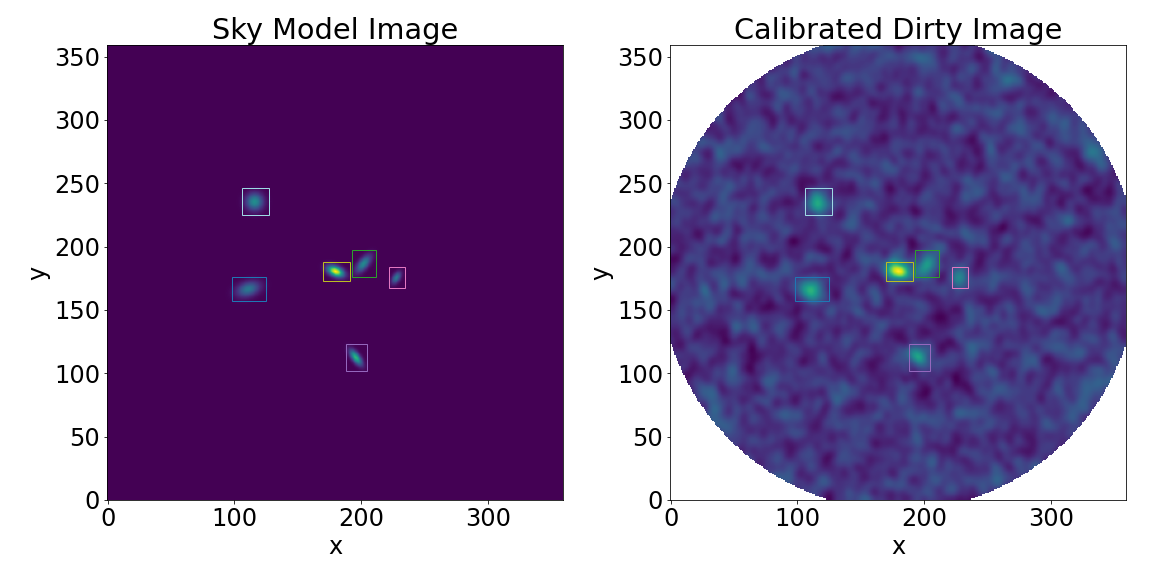}
	\end{subfigure}

	\begin{subfigure}[b]{1\columnwidth}
		\includegraphics[width=\textwidth]{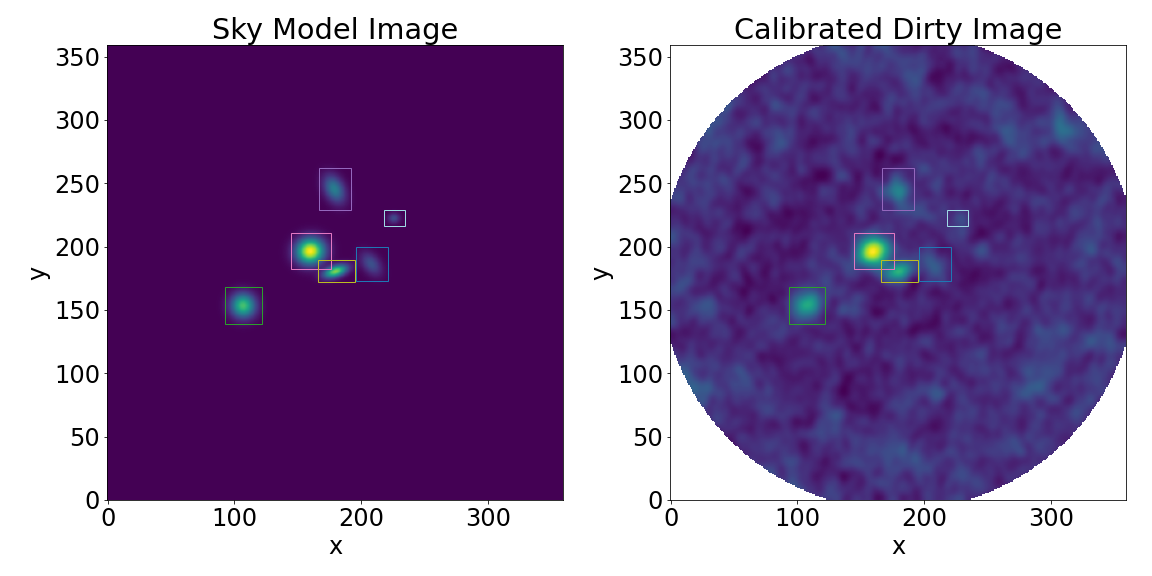}
	\end{subfigure}
	\hfill
	\begin{subfigure}[b]{1\columnwidth}
		\includegraphics[width=\textwidth]{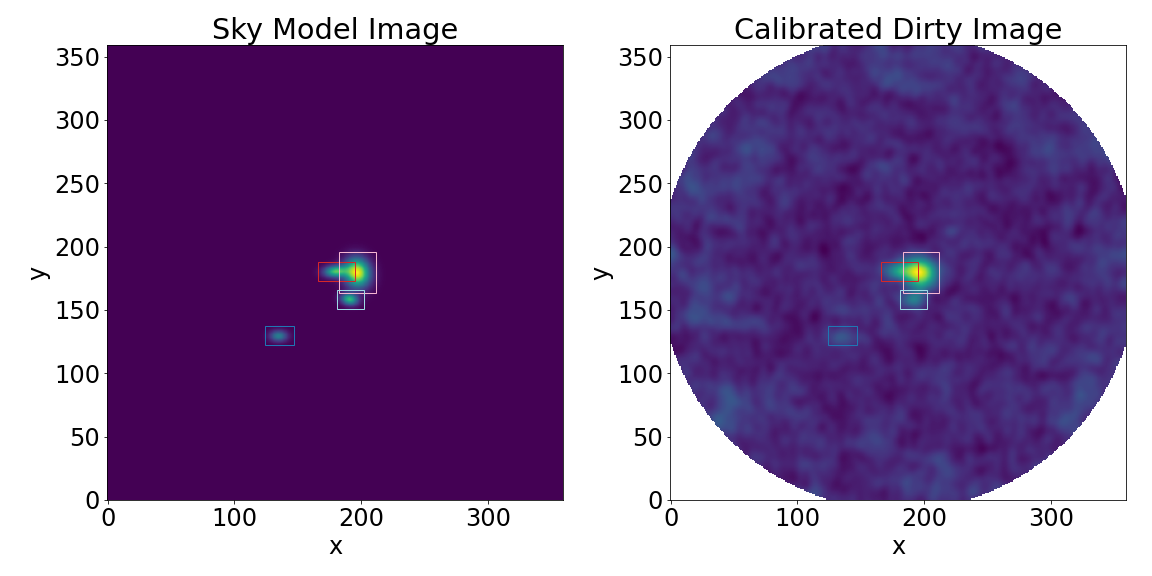}
	\end{subfigure}
	
	\caption{Several examples of frequency stacked dirty/clean cube pairs generated through our simulation code.
			Sources within the cubes are outlined with colored bounding boxes.}
	\label{fig:sim_examples}
\end{figure*}

We use the ALMA Cycle 9 C-3 for the antenna configuration, simulating $43$ antennas
in the 12-m Array with a maximum baseline of $0.50$ km (\href{https://almascience.eso.org/documents-and-tools/cycle9/alma-technical-handbook}{ALMA Cycle 9 Technical Handbook}). Once the measurement sets have been produced, 
to create the dirty cubes, we feed them to the \textsc{TCLEAN} task initialized with the parameter \textit{niter}
set to zero. This forces the task not to run the \textsc{TCLEAN} algorithm but only to perform the fast 
Fourier transform of the visibility data and  the gridding using the pixel parameters outlined in 
Tab~\ref{tab:simobserve_parameters}. The resulting data products are noisy cubes
convolved with the ALMA synthesized beam with an angular size of $10$ squared arcseconds and a total bandwidth of $1.28$ GHz.
The spatial dimensions of the cubes are $360 \times 360$ pixels and they are characterised by $128$ channels in frequency. 
Nevertheless, the simulations are already using Pardo's ATM library to construct an atmospheric profile 
for the ALMA site comprising a water layer in the atmosphere and thermal noise. After the dirty cubes are produced, 
we inject 3D uncorrelated white Gaussian noise. The RMS of the noise is automatically adjusted to 
reach a target SNR measurement for the central source. This way, we can produce simulations containing 
sources with specific SNRs, which helps testing the pipeline capabilities and its detection limits. 
After the dirty cubes are produced, the morphological parameters, stored in the .csv parameter file,
are used to measure further source properties of interest, such as the continuum, the SNR, and the total surface brightness.
The sky model and dirty cubes can be produced both sequentially or in parallel through the \textit{Slurm} workload manager \citep{slurm} if a multi-node architecture is available.
Fig.~\ref{fig:sim_examples} shows several examples of frequency stacked clean and dirty cubes pairs.
While we plan to improve our simulation code beyond its current capabilities, at this stage it does not model galaxies internal kinematics, it does not include complex galaxy morphologies and it does not account for galaxy interactions. That being said, given the main scientific target of the present work, these parameters should not play an important role in shaping the real target morphologies given that they are mostly unresolved. For this reason, we generate sources with a 2D Spatial Gaussian with a FWHM between 0.2 and 0.8 arcsec. Regarding the simulation of complex atmospherical and instrumental effects, the \textsc{CLEAN} \textit{SimALMA} pipeline, that we employ to simulate ALMA observations of the sky models, should produce fairly realistic ALMA observations in case of point-like sources \citep{CASA2022}.
While it is out of the scope of this work to present the full simulation pipeline, it is our intention, given the relevance of comparing the performances of deep learning models on the same data, to make it publicly available on GitHub with several baseline datasets, such as the one we used to train and test our pipeline.

\subsection{The Data}\label{sec:data}
\begin{figure*}
	\centering
	\begin{subfigure}[b]{1\columnwidth}
		\includegraphics[width=\textwidth]{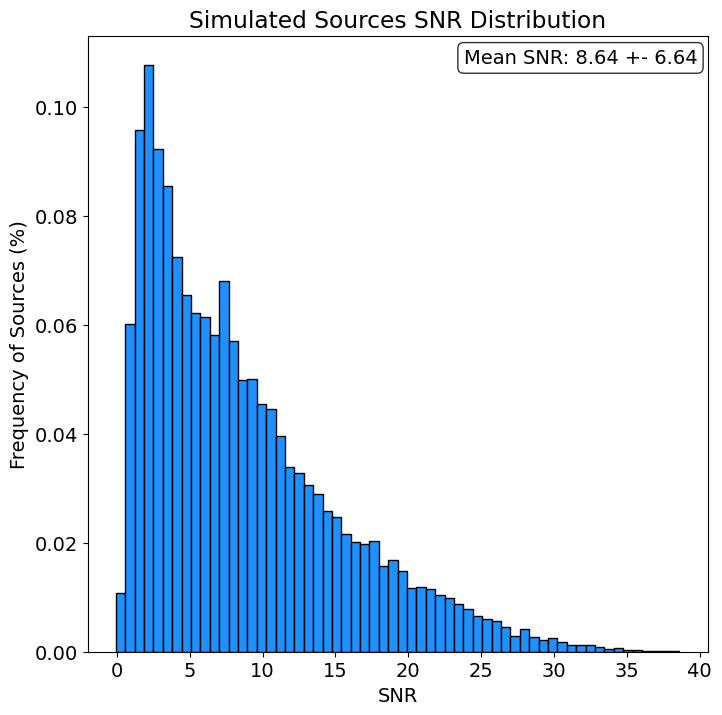}
		\subcaption{Distribution of the Signal to Noise Ratio of the simulated sources: fraction of simulated sources versus measured 
		SNR (see Eq.~\ref{eq:global_SNR}). The box in the top right
		corner shows the mean SNR $\pm$ its standard deviation. }
	\end{subfigure}
	\hfill
	\begin{subfigure}[b]{1\columnwidth}
		\includegraphics[width=\textwidth]{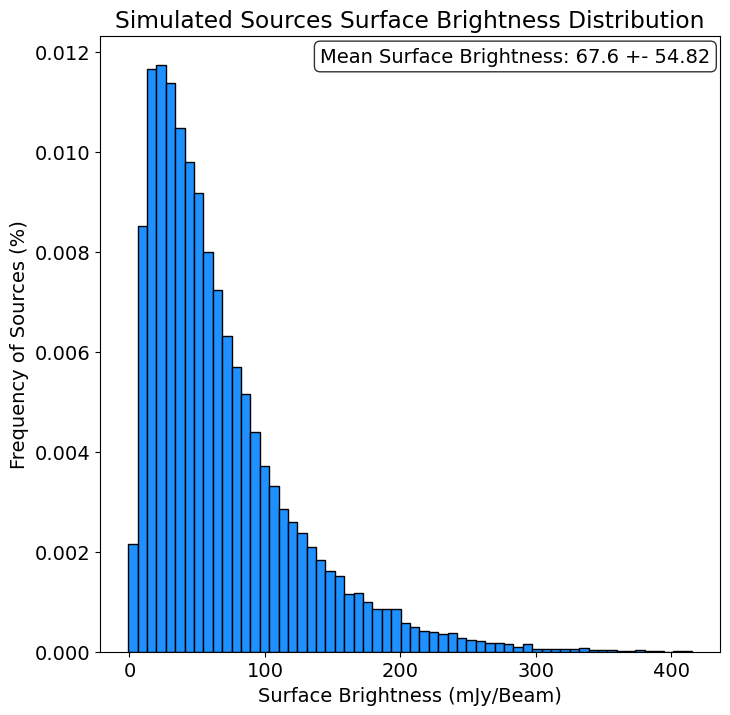}
		\subcaption{Distribution of the the total brightness of the simulated sources. On the x-axis, the
		measured total brightness [mJy / beam] is obtained by summing the voxel values in the dirty cubes within 
		sources bounding boxes. On the y-axis, the fraction of simulated sources is provided. The box in the top right corner shows
		the mean brightness $\pm$ its standard deviation.}
	\end{subfigure}

	\begin{subfigure}[b]{1\columnwidth}
		\includegraphics[width=\textwidth]{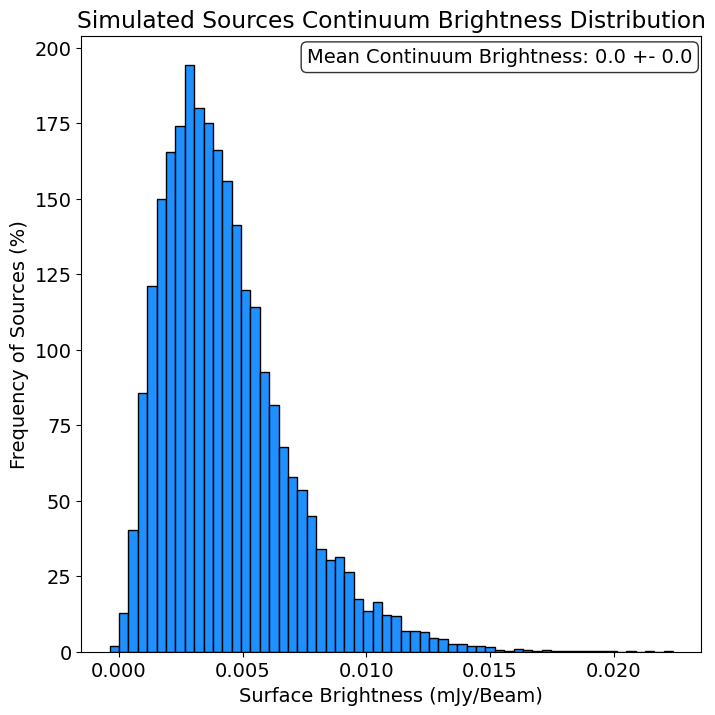}
		\subcaption{Distribution of the continuum mean brightness of the simulated sources. On the x-axis, the
		measured mean continuum brightness [mJy / beam] is obtained by selecting all voxels within the $x$ and $y$ limits of the sources
		bounding boxes but outside their boundaries in frequency $[z - fwhm_z, z + fwhm_z]$. On the y-axis, the fraction of
		simulated sources is shown. The box in the top right corner provides the mean continuum brightness $\pm$ its standard deviation.}
	\end{subfigure}
	\hfill
	\begin{subfigure}[b]{1\columnwidth}
		\includegraphics[width=\textwidth]{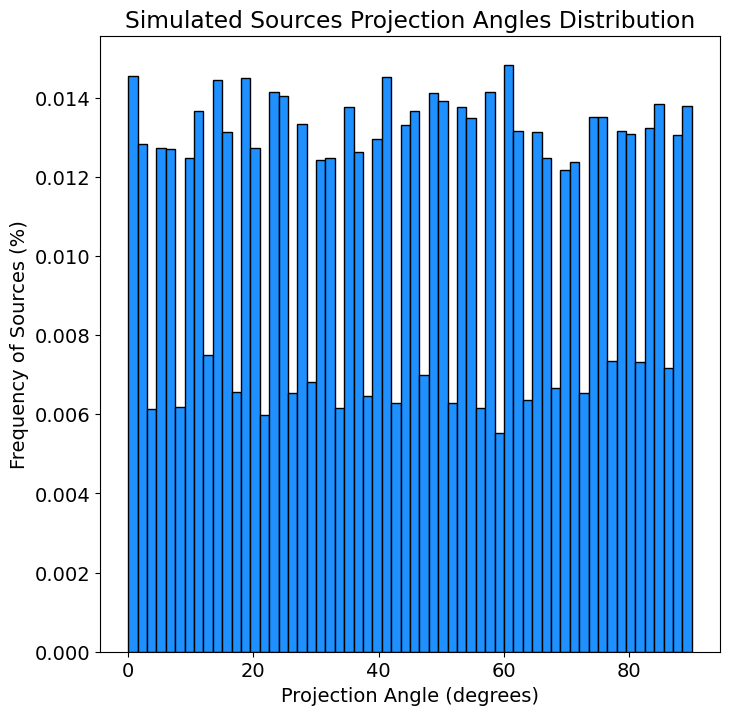}
		\subcaption{Distribution of the projection angles of the simulated sources: fraction of simulated sources versus projection 
		angles in degrees}
	\end{subfigure}
	\caption{The figure shows (a) the distribution of sources SNRs, (b) the distribution of sources surface brightness, 
	(c) the distribution of sources mean continuum brightness, and (d) the distribution of sources projection angles.}
	\label{fig:simulated_parameters_1}
\end{figure*}

\begin{figure*}
	\centering
	\begin{subfigure}[b]{1\columnwidth}
		\includegraphics[width=\textwidth]{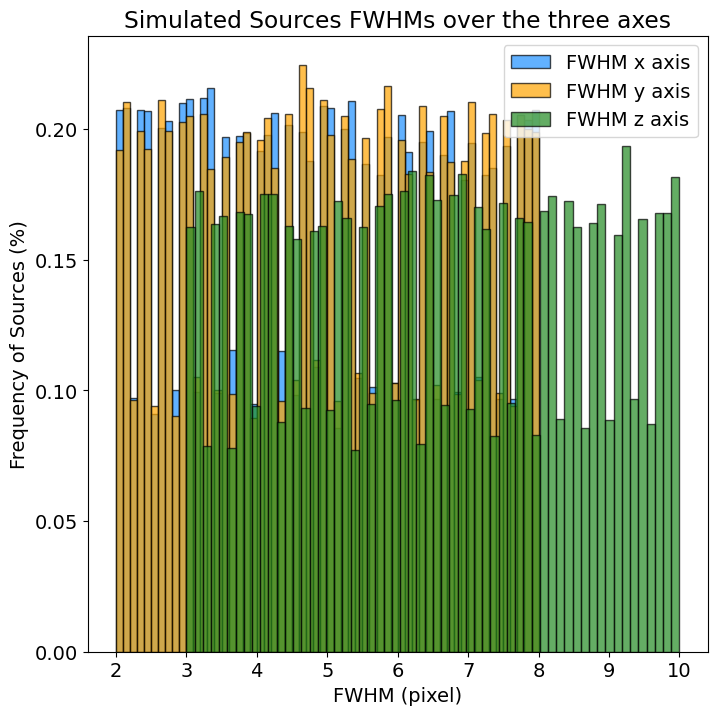}
		\subcaption{Distributions of the $FWHMs$ of the simulated sources. In blue, orange and green, the $FWHMs$ over the x-axis,  the y-axis and the z-axis are provided. On the histogram, the x-axis and y-axis provide 
		the $FWHMs$ values in pixels and the fraction of simulated sources, respectively.} 
	\end{subfigure}
	\hfill
	\begin{subfigure}[b]{1\columnwidth}
		\includegraphics[width=\textwidth]{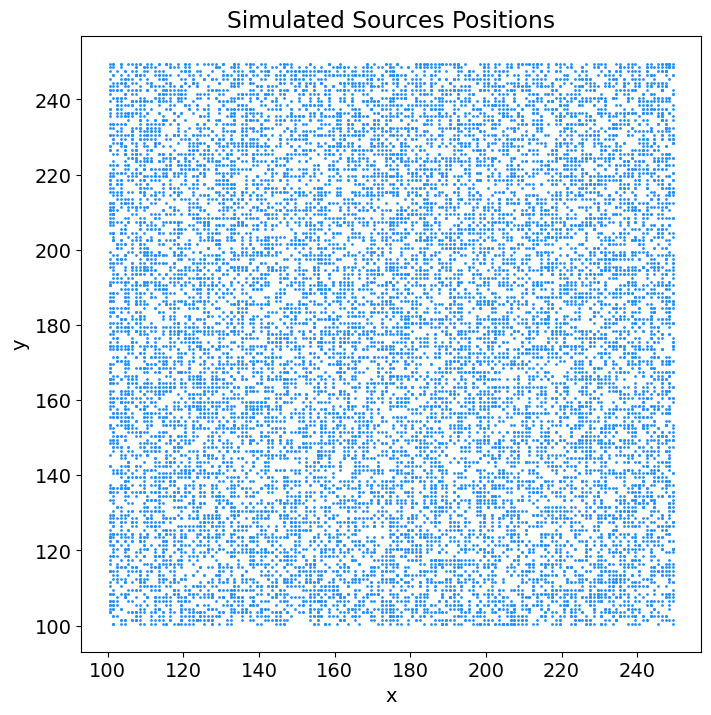}
		\subcaption{Scatter plot showing uniformity in the positions on the $xy$ plane of the simulated sources.}
	\end{subfigure}
	\caption{The figure shows (a) the distribution of the $FWHMs$ of the sources over the three cube axes, and (b)
	the distribution of the sources positions over the $xy$ plane.}
	\label{fig:simulated_parameters_2}
\end{figure*}

We generated $5,000$ simulated cube pairs containing $22,532$ simulated sources and randomly divided them into train, validation, and test sets using the rather usual 
60\%, 20\%, 20\% splitting criterium.
We just remind the reader that the training set is used to train the DL models within the pipeline; the validation set is used to  measure the training progress and assess generalization capabilities, and the test set is used to measure the pipeline performances in detecting sources and in regressing their parameters (see Sec.~\ref{subsec:pipeline} and \ref{subsec:training}). 
The three sets contain respectively $13,512$, $4,465$, and $4,556$ simulated sources.
 The distributions of the source parameters are shown in Fig~\ref{fig:simulated_parameters_1} and Fig~\ref{fig:simulated_parameters_2}.
Projection angles, positions, and extensions of sources are uniformly generated in their 
 respective parameter ranges (see Tab.~\ref{tab:simobserve_parameters}), while the SNR, Surface Brightness, and Continuum Brightness distributions
 confirm that we are generating a bright central source with a $SNR > 10$ surrounded by less bright serendipitous sources. The minimum and maximum flux densities generated are respectively $0.97$ and $407.4$ mJy/beam.
The data were generated on the IBISCO-HPC (Infrastructure for Big data and Scientific Computing) at the University of Naples Federico II (\href{https://ibiscohpc-wiki.scope.unina.it/wiki:startdoc}{IBISCO-HPC}). 
\textbf{}

\section{The Pipeline}\label{sec:dl}
The Deep Learning Pipeline we present in this work, as it will be more apparent in Sec.~\ref{subsec:pipeline}, can be described as a decision
graph interconnecting six deep learning models, each one taking a specialised role in order to detect and characterise sources within the input dirty cubes. The types of architectures were chosen on the basis of their strengths: Convolutional architectures (Blobs Finder and ResNets) to process spatial information, and the Recurrent Neural Network (Deep GRU) to process sequential information. 
Before describing the flow of data within the pipeline, we shortly describe the individual DL models. Note that all our models were implemented through the PyTorch library \citep{pytorch}.

\subsection{Blobs Finder}\label{subsec:blobsfinder}
\begin{figure*}
	\centering
	\includegraphics[width=0.8\textwidth]{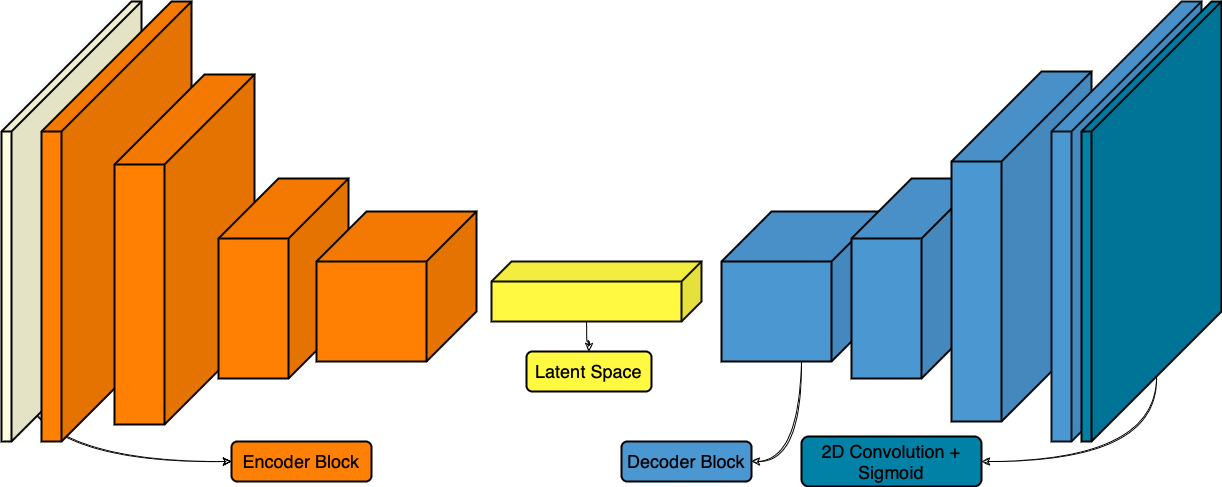}
	\caption{Blobs Finder's architecture. The Deep Convolutional Autoencoder
		is made by an encoder and a decoder networks. The Encoder is made by four convolutional
		blocks each one constituted by a 2D Convolution layer with stride $2$ and a kernel size of $3$, a Leaky 
		ReLU activation function and a 2D Batch Normalization layer followed by a 2D Convolution layer
		with a stride of $1$ and a kernel size of $3$ a Leaky ReLU activation function and a 2D Batch Normalization Layer. 
		The final layer
		is fully connected layer with an output size of $1024$. The Decoder is made by $4$
		deconvolutional blocks and a final block. The convolutional blocks are
		constituted by a 2D Bilinear interpolation with stride of $2$, followed by a 
		Leaky ReLU activation function and a 2D Batch Normalization layer (upsampling block),
		a 2D Transposed Convolution layer with stride of $2$ and a kernel size of $3$,
		followed by a Leaky ReLU activation function and a 2D Batch Normalization layer (learnable upsampling block). 
		The output of the up-sampling block and learnable up-sampling block are then concatenated and passed along to a convolutional block constituted by a 2D Convolution layer with stride $2$ and a kernel size of $3$,
		a Leaky ReLU activation function and a 2D Batch Normalization layer followed by a 2D Convolution layer
		with a stride of $1$ and a kernel size of $3$. The final block is
		a 2D Convolution layer with a stride of 1 and a kernel size of 1, followed by a
		Sigmoid activation function.}
	\label{fig:blobsfinder}
\end{figure*}
Blobs Finder is a 2D Deep Convolutional Autoencoder trained to solve
the image deconvolution problem
\begin{equation}
	D[x, y] = P[x, y] \times M[x, y] + N[x, y]
\end{equation}
where $D[x, y]$ is the integrated dirty cube produced integrating along the frequency the dirty cube, $P[x, y]$ is the dirty PSF, and $N[x, y]$ is the combination of
all noise patterns in the data. $M[x,y]$ is the reconstructed and denoised integrated sky image.
The idea behind an Autoencoder \citep{autoencoder} is that relevant features in the data, such as the shape of the structures within the image, should be relatively
robust with respect to the noise components, and hence can be learned
by compressing the data to a lower dimensional latent space. The latent space representation can then be used to reconstruct a \textit{noiseless} version of the input data.
To guide the autoencoder towards the best possible solution, two factors are critical \citep{Goodfellow-et-al-2016}: the preprocessing and augmentation of input and target variables, and the choice of the loss function used to measure the error between the target variable and the Autoencoder's prediction. 
In our case, Blobs Finder is trained with the integrated dirty cubes as inputs, and the sky model images as targets. Both input and target image are normalized to the $[0, 1]$ range, which helps with the training process and allows us to make a probabilistic interpretation of the Autoencoder's output.
Blobs Finder architecture, as shown in Fig.~\ref{fig:blobsfinder}, is that of a Convolutional Encoder \citep{convolutional_autoencoder} structured by four convolutional blocks that progressively reduce the spatial dimension of the input while increasing the number of channels or feature maps. More in detail, each block contains a 2D Convolution layer with stride $2$ and a kernel size of $3$, a Leaky ReLU (Rectified Linear Unit) activation function and a 2D Batch Normalization layer followed by a 2D Convolution layer with a stride of $1$ and a kernel size of $3$, another Leaky ReLU activation and a 2D Batch Normalization Layer. In this way, each block halves the spatial extent of the input and doubles the number of channels. After the convolutional blocks there is the final, fully connected layer. 

The Decoder has a symmetric architecture, i.e. a fully connected layer, followed by four deconvolutional blocks and a final identity layer which is constituted by a 2D Convolution followed by a Sigmoid activation function.
Each deconvolutional block is constituted by a 2D Bilinear interpolation with stride of $2$, followed by a  Leaky ReLU activation function and a 2D Batch Normalization layer (upsampling block), a 2D Transposed Convolution layer with stride of $2$ and a kernel size of $3$,
followed by a Leaky ReLU activation function and a 2D Batch Normalization layer (learnable upsampling block). 
The output of the upsampling block and learnable upsampling block are then concatenated and passed along to a convolutional blocks constituted by a 2D Convolution layer with stride $2$ and a kernel size of $3$, a Leaky ReLU activation function and a 2D Batch Normalization layer followed by a 2D Convolution layer with a stride of $1$ and a kernel size of $3$. 

The bilinear upsampling operation transform the input layer in the desired spatial resolution without using any parameters and the resulting features should contain most of the information of the original features. These upsampled features are concatenated to the output of the parametric upsampling performed through the Transposed Convolution in order to create a residual-like connection. This additive upsampling \citep{wojna} should improve prediction capabilities and remove gridding artefacts produced by the subsequent Transposed Convolutions.
The final block is a 2D Convolution layer with a stride of 1 and a kernel size of 1, followed by a Sigmoid activation function. The scope of this final layer is to normalize its input to the $[0, 1]$ range. A detailed description of the spatial transformation performed by Blobs Finder can be seen in Tab.~\ref{tab:autoencoder_shape}.
\begin{table}
	\centering
	
	\begin{tabular}{lll}
		\hline
		Block Name     & Input Size                    & Output Size                   \\
		\hline
		Conv Block 1   & $[b, 1, 256, 256]$            & $[b, 8, 128, 128]$            \\
		Conv Block 2   & $[b, 8, 128, 128]$            & $[b, 16, 64, 64]$             \\
		Conv Block 3   & $[b, 16, 64, 64]$             & $[b, 32, 32, 32]$             \\
		Conv Block 4   & $[b, 32, 32, 32]$             & $[b, 64, 16, 16]$             \\
		FC 1           & $[b, 64 \times 16 \times 16]$ & $[b, 1024]$                   \\
		\hline
		FC 2           & $[b, 1024]$                   & $[b, 64 \times 16 \times 16]$ \\
		DeConv Block 1 & $[b, 64, 16, 16]$             & $[b, 32, 32, 32]$             \\
		DeConv Block 2 & $[b, 32, 32, 32]$             & $[b, 16, 64, 64]$             \\
		DeConv Block 3 & $[b, 16, 64, 64]$             & $[b, 8, 128, 128]$            \\
		DeConv Block 4 & $[b, 8, 128, 128]$            & $[b, 1, 256, 256]$            \\
		Final Block    & $[b, 1, 256, 256]$            & $[b, 1, 256, 256]$            \\
		\hline
	\end{tabular}
	\caption{Input and Output shapes for each layer of Blobs Finder, where $b$ indicates the batch size, and the horizontal line separates the Encoder from the Decoder network.}
	\label{tab:autoencoder_shape}
\end{table}
To train Blobs Finder, we selected as a loss function the weighted combination of two
well-known losses in the DL image reconstruction and denoising framework: the $l_1$ loss and the
Structural dissimilarity loss $DSSIM$ which is based on Structural Similarity Index measurement.
The losses are mathematically defined as follows:
\begin{equation}
	\label{eq:l1}
	l_1(x, y) = MAE(x, y) = mean([l_1, ......, l_N]^T)
\end{equation}

where $N$ is the number of pixels in the images and
\begin{align}
	DSSIM(x, y) = \frac{1 - SSIM(x, y)}{2}                                                                                   \\
	SSIM(x, y) = \frac{(2 \mu_x \mu_y + c_1)(2 \sigma_{xy} + c_2)}{(\mu_x^2 + \mu_y^2 + c_1)(\sigma_x^2 + \sigma_y^2 + c_2)} \\
	c_1 =(k_1 L)^2                                                                                                           \\
	c_2 = (k_2 L)^2                                                                                                          \\  
	L = 2^{precision} - 1
\end{align}
where $\mu_x$ and $\mu_y$ are the averages of $x$ and $y$, $\sigma_x^2$ and $\sigma_y^2$ are their variances, $\sigma_{xy}$ the covariance, $c_1$ and $c_2$ two variables needed to stabilize the division with small denominator values, and, finally, $L$ is the dynamic range of the pixel values. 
In our case, the images are stored in single-precision floating-point format and thus $L = 2^{32} -1$ and the two constants are respectively $k_1 = 0.01$ and $k_2 = 0.03$.

The choice of these two losses was empirically determined by trial and error in our experiments. More information about how these losses are used and their weighting during training is outlined in Sec.~\ref{subsec:training}.

\subsection{Deep Gated Recurrent Unit (GRU)}\label{subsec:gru}
\begin{figure}
	\centering
	\includegraphics[width=\columnwidth]{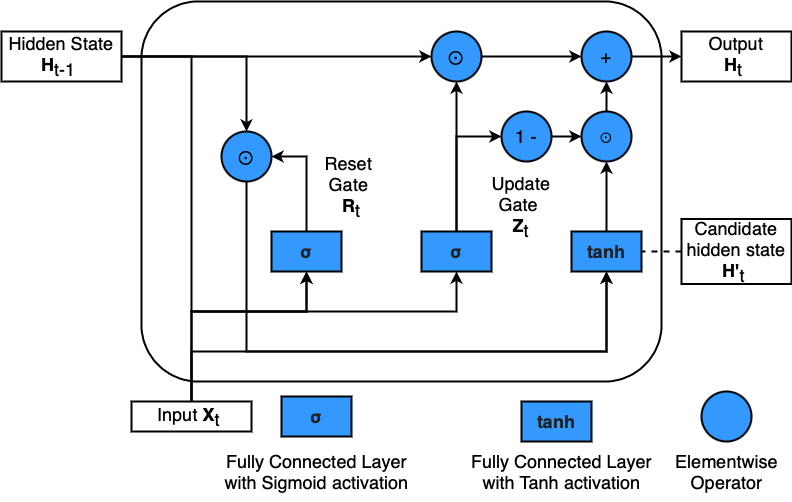}
	\caption{Gated Recurrent Unit architecture showing the flow of data within the
		network.}
	\label{fig:gru}
\end{figure}

\begin{figure}
	\centering
	\includegraphics[width=\columnwidth]{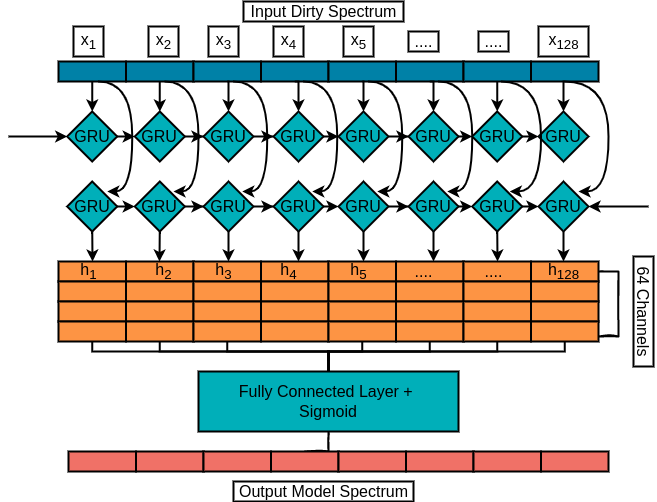}
	\caption{Deep GRU's architecture constituted by two layers of GRUs followed by
		a Fully connected layer and a Sigmoid activation function.}
	\label{fig:deep-gru}
\end{figure}

Deep Gated Recurrent Unit (GRU) is a Recurrent Neural Network (RNN) \citep{RNN} constructed by combining together two layers of GRUs and a fully connected layer, as shown in Fig.~\ref{fig:deep-gru}.
RNNs were designed in order to capture correlation in sequences of data (usually 1D signals).  GRU \citep{GRU} tries to solve  the main shortcoming of RNNs (namely the exploding or vanishing gradients problem) by introducing gating of its hidden states. 
Gating is introduced through the \textit{reset} and \textit{update} gates which, respectively, control how much of the previous hidden state must be remembered and the degree to which the current hidden state is similar to the previous one at each frequency iteration. The first helps capturing short-term correlations in the data, while the second
 one captures the long-term correlations.
As shown in Fig.~\ref{fig:gru}, which outlines the data flow inside a GRU, these gates are  implemented through fully connected layers with a sigmoid activation function:
\begin{align}
	R_t = \sigma(X_t W_{xr} + H_{t - 1} W_{hr} + b_r) \\
	Z_t = \sigma(X_t W_{xz} + H_{t - 1} W_{hz} + b_z)
\end{align}
where $X_t$ is the input vector, $H_t$ is the hidden state at the previous frequency step,
and $W$ and $b$ are the weights and biases of the fully connected layers.

The reset gate is multiplied (element-wise) with the previous hidden state and the effect of this multiplication is passed along to the candidate hidden state
\begin{equation}
	H'_t = tanh(X_t W_{ht} + (R_t \odot H_{t - 1}) W_{hh} + b_h)
\end{equation}
where $W_{ht}$ and $W_{hh}$ are the model weights, $b_h$ is the bias term and $\odot$ is the Hadamard element-wise product.

If the entries of $R_t$ are close to $1$, the GRU behaves just as simple RNN, while if it is close to $0$, the previous hidden state is forgotten and the RNN behaves like a Multi Layer Perceptron (MLP).
The final hidden state, is given by the element-wise convex combination of $H_{t - 1}$
and $H'_t$
\begin{equation}
	H_t = Z_t \odot H_{t - 1} + (1 - Z_t) \odot H'_t
\end{equation}
If the entries of $Z_t$ are close to $1$, the hidden state at the previous iteration is retained and the information brought by the current input $X_t$ is ignored. Otherwise, if the entries of $Z_t$ are close to $0$, the old hidden state is updated with the new candidate hidden state $H'_t$.
In our case, we want to use the Deep GRU to solve the denoising problem
\begin{equation}
	Y[z] = X[z] + N[z] \quad  \textrm{with} \ z \in [1, 128]
\end{equation}
where $z$ is the frequency index in the cube, $Y[z]$ are noisy galaxy spectra, $X[z]$ are the underlying emissions, and $N[z]$ are the various noise components. When a spectrum is fed to the network, each hidden state at frequency step $t$ is passed to both the next frequency step
of the current layer and the current frequency step of the next layer. 
Because there is no preferable frequency direction in the data, the two layers pass along information in opposite directions:
one from low frequencies to high frequencies, and the other in the opposite direction.
In our implementation, each layer of GRUs outputs 32 hidden states (or channels)
which are then concatenated to form a latent vector of size $[b, 64 \times 128]$ before
being fed to a  fully connected layer. 
The layer transforms its input in a vector with the same size as the input signal
and then a Sigmoid activation function is applied to normalize it to the $[0, 1]$ range.
As loss function, we use again the $l_1$ loss (see Eq.~\ref{eq:l1}).

\subsection{ResNet}\label{subsec:resnet}
\begin{figure}
	\centering
	\includegraphics[width=\columnwidth]{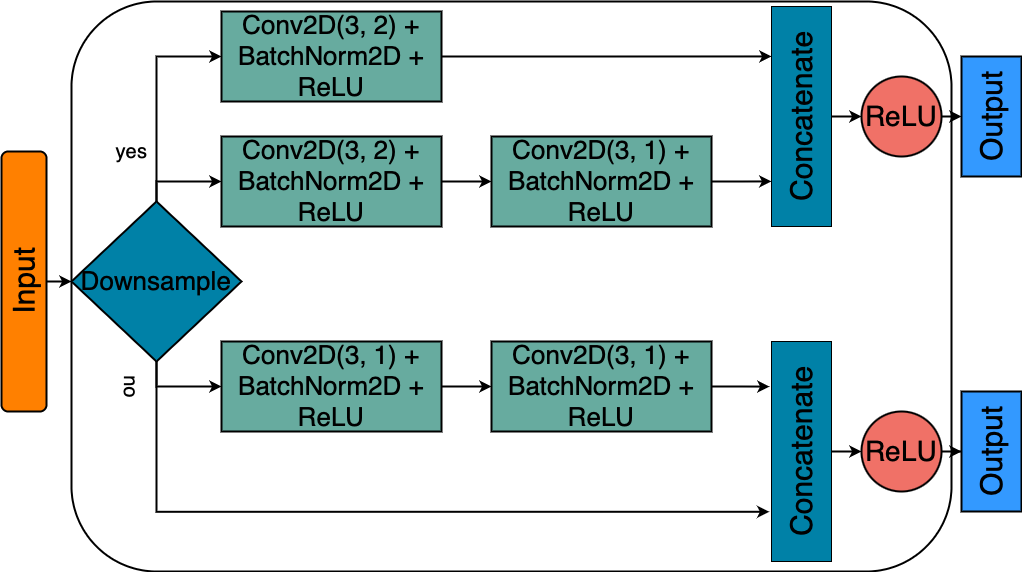}
	\caption{Architecture of a Residual Block, the basic component of a ResNet.
		The architecture is divided into two main pathways depending if downsampling must be applied
		in the layer. In the affirmative case, the 2D Convolutions
		are applied with a kernel size of $3$ and with a stride of $2$ or $1$. As it can be seen
		the output of the previous layer is brought forward through a skip-connection and concatenated
		with the output of the current layer before applying the final activation function.}
	\label{fig:resblock}
\end{figure}

Residual Neural Networks \citep{ResNet} are a class of Deep Convolutional Neural Networks (CNNs) which implements skip connections between adjacent layers in order to avoid the problem of vanishing gradients \citep{vanishing-gradient} by easing the process of information flow in the layers and to mitigate the degradation problem, a phenomenon for which the more layers are added to a CNN the more training error and instability increase. The basic component of a ResNet is the so called Residual Block outlined in Fig.~\ref{fig:resblock}.

Our implementation of the ResBlock depends on whether spatial downsampling is applied
to the input or not. 
If spatial downsampling is needed, the input is processed via two convolutional blocks
constituted by a 2D Convolutional Layer with kernel size of 3 and a stride of 2 (which downsamples the input), a ReLU activation function and a 2D Batch Normalization layer. While the first convolutional block increases the number of channels by a factor of two, the second maintains constant the number of channels. In parallel, the input is also processed by a single
convolutional block, and the two outputs are concatenated together (skip-connection) before being fed to a final ReLU layer. 
On the other case, if downsampling is not necessary, the input is processed
via two convolutional blocks with a stride of $1$ (which thus does not downsample the input) and is concatenated with the unaltered input (skip-connection) before being fed to the final ReLU.
\begin{figure*}
	\centering
	\includegraphics[width=\textwidth]{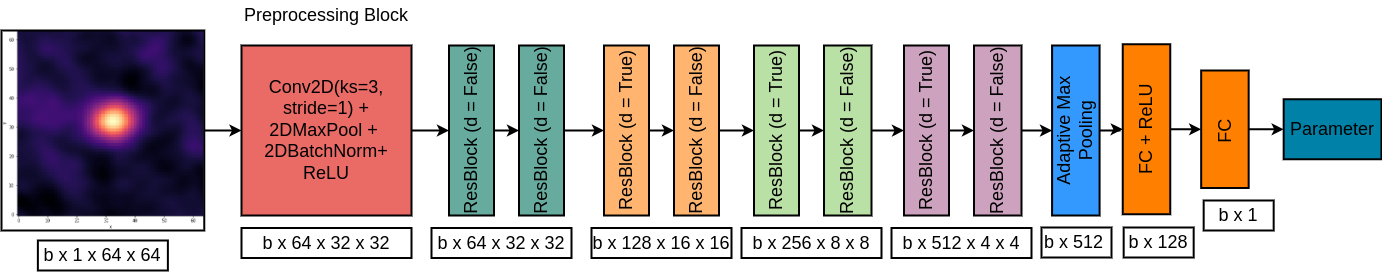}
	\caption{Our implementation of the ResNet architecture. The input image is first preprocessed
		by a 2D Convolution layer, followed by 2D Max Pooling, 2D Batch Normalization, and a ReLU activation
		function, and then is forwarded through four blocks of two Residual Blocks (see Fig.~\ref{fig:resblock}). The output is then processed
		via an Adaptive Max Pooling layer and fed to two fully connected layers which map the latent vector of
		$512$ elements to a single scalar (the value of the parameter of interest for the ResNet).}
	\label{fig:resnet}
\end{figure*}

The ResNet architecture, which is shown in Fig.~\ref{fig:resnet}, is constructed by a first
convolutional block that performs 2D Max Pooling followed by four ResBlocks which gradually
spatially downsample the input image of size $[b, 1, 64, 64]$ to $[b, 512, 4, 4]$ where $64 \times 64$ are the spatial dimensions of the intput image, $b$ is the batch size, $1$ is the initial number of channels and $512$ is the number of feature maps which are fed to the final MLP. Average Max Pooling
is applied to collapse the spatial dimensions to a single vector which is then fed to two fully connected layers interconnected by a ReLU activation function (MLP). This final layer outputs a single scalar value.
In other words, the first part of the network extrapolates a vector of $512$ features from the input image, which is then fed to a Multi Layer Perceptron (MLP) that makes the functional mapping between the features space and the target parameter space.
In our pipeline, we use several ResNets; each one specialized in solving the regression of a specific morphological parameters of the input source. Technically, these parameters could be regressed at the same time by a single ResNets with an output vector of size $m$ instead of a scalar (where $m$ is the number of parameters one wants to regress). Given that the dynamical range of the parameters varies a lot from one parameter to the other (see Sec.~\ref{sec:data}), we expect that such a general network would show lower performances with respect to the specialized counterparts. This expectation was empirically confirmed by us on a trial and error base. 
As loss function to train the ResNets, we used again the $l_1$ loss (see Eq.~\ref{eq:l1}).

\subsection{The Pipeline} \label{subsec:pipeline}
\begin{figure*}
	\centering
	\includegraphics[width=\textwidth]{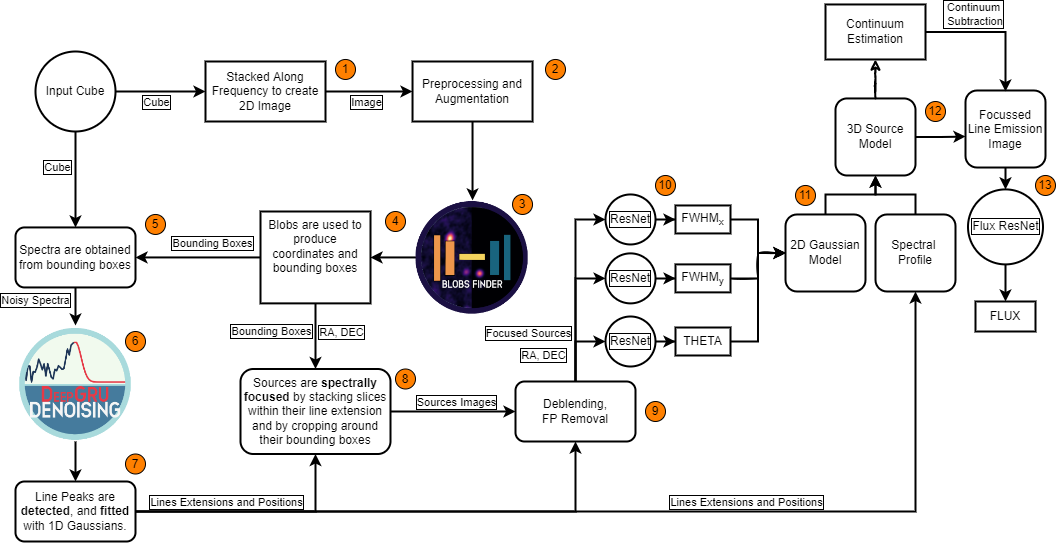}
	\caption{The full pipeline schema. Numbers show the logical flow of the
		data within the pipeline.}
	\label{fig:pipeline}
\end{figure*}

The overall objective of the pipeline is to detect and fit the sources within calibrated ALMA image cubes. A full overview of the pipeline can be seen in Fig.~\ref{fig:pipeline}, where the
arrows outline the flow of data in the pipeline. The order in which the operations
are performed can be understood by the numbering (in orange).
As already explained, the pipeline can be divided into six logical blocks: 2D source detection, frequency denoising, and emission detection, source focusing, morphological parameters estimation, 3D model construction and flux density estimation.
To relate the logical blocks to the pipeline schema shown in Fig.~\ref{fig:pipeline},
in the following explanation, we mark the logical blocks with the corresponding
numbers as shown in Fig.~\ref{fig:pipeline}. 
To ease the logical flow of the pipeline, we assume that all the DL models have been trained to
act as simple, functional maps between their inputs and outputs. 
In Sec ~\ref{subsec:training} we go over the details of how each Deep Learning model is trained within the pipeline. All the images shown in this section are relative to the Test set; more examples and a detailed description of the pipeline performances
are described in Sec.~\ref{sec:source_detection}.

The flow of an input dirty cube in the pipeline goes as it follows:
\begin{enumerate}
 	\item \textbf{2D Sources Detection} (\textbf{1 - 4}): the image cube is normalized to the $[0, 1]$ range and  integrated along frequency to create a
	      2D image. We refer to this image as the \textbf{integrated dirty cube} (1). The integrated dirty cube
	      is first cropped from the centre to a shape of $[256, 256]$ pixels which removes the edge of the images characterized by a low SNR (See Sec.~\ref{sec:simulations}), then it is
	      normalized to the $[0, 1]$ range (2) and, finally,  it is
	      fed to the first DL model \textbf{Blobs Finder}  (Sec.~\ref{subsec:blobsfinder}).
	      The autoencoder
	      processes the image and predicts a \textbf{2D probabilistic map} of source detection (3).
	      A hard thresholding value of $0.15$
	      is used to binarize the probabilistic map and then the \textit{scikit-learn}
	      \citep{scikit-learn}
	      \textit{label} and \textit{regionprops} functions are used to extract bounding
	      boxes around all blobs of connected pixels (or source candidates) (4). The thresholding value is chosen to be $0.15$ in order to peak all the signal detected by Blobs Finder, while excluding small fluctuation in the background.
	      The exact thresholding value was empirically determined in order to strike a compromise between the number of false detections and the percentage of the detected signal. 
 Figures~\ref{fig:dirty_map}, \ref{fig:boundingboxes}, and \ref{fig:target_map}
		  show, respectively, an example of an input integrated dirty cube containing $6$ simulated sources 
		  (outlined by green bounding boxes and two of which spatially blended), the target sky model image 
		  (with in green the target bounding boxes and in red the predicted bounding boxes extracted through the thresholding of the 
		  predicted 2D probabilistic map), and the 2D prediction map with  the predicted bounding boxes highlighted in red.
    
	      \begin{figure}
		      \centering
		      \includegraphics[width=\columnwidth]{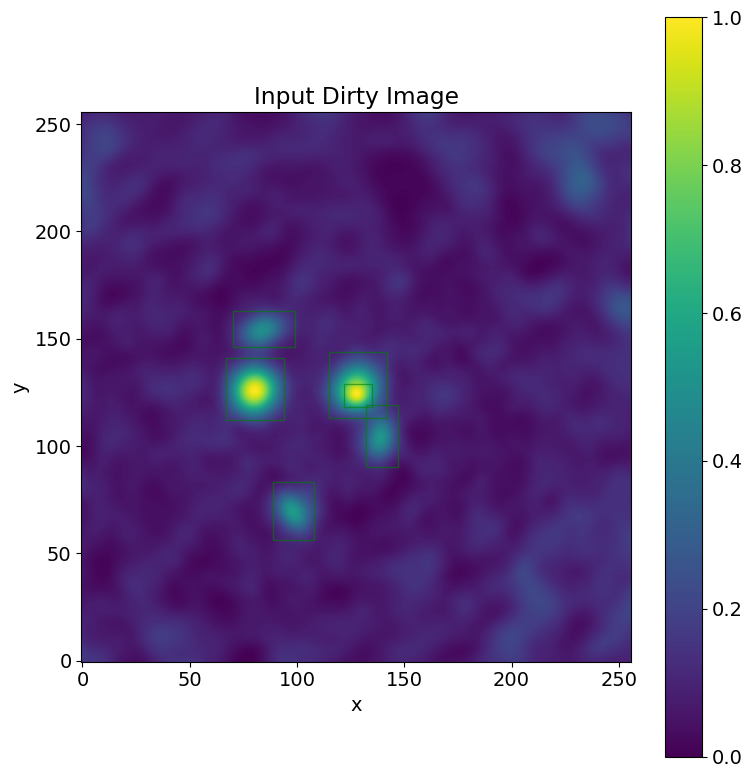}
		      \caption{An example of Blobs Finder's input 2D integrated dirty cube
			      produced by integrating an input dirty cube over the entire
			      frequency range. Superimposed in green, are the target bounding boxes outlining the emissions 
				  of the $6$ sources present in the cube. The image contains an example of two spatially blended 
				  sources located around the centre of the image, one is a bright point-like source, the other a fainter and diffuse source laying behind.}
			\label{fig:dirty_map}  
	      \end{figure}

		  \begin{figure}
			\centering
			\includegraphics[width=\columnwidth]{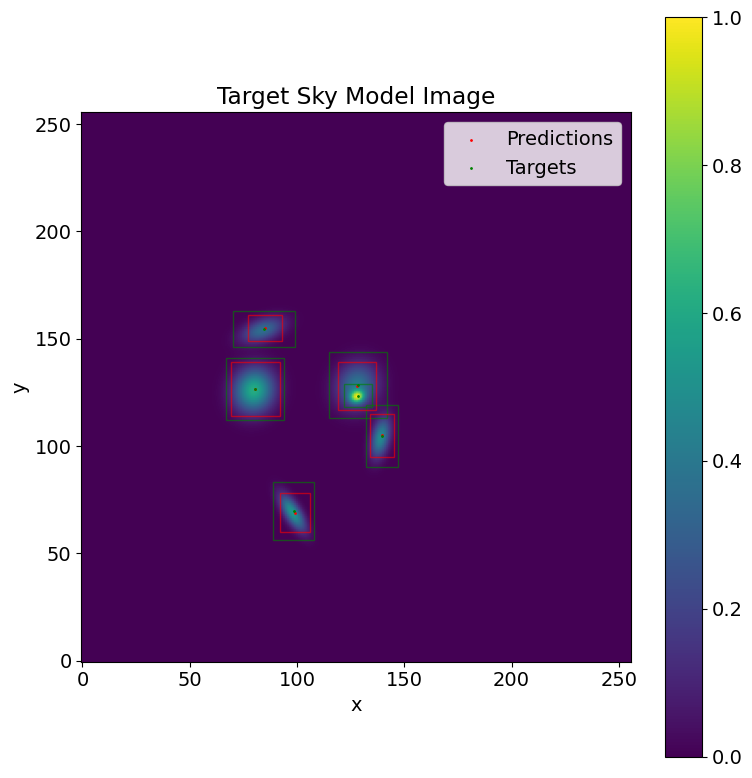}
			\caption{An example of Blobs Finder's target 2D Sky Model image
				with the target bounding boxes highlighted in green and the predicted bounding boxes
				extracted through the tresholding operation on Blobs Finder's probabilistic output, 
				highlighted in red. Predicted and true bounding box centers are also plotted as, respectively, 
				red and green dots.}
			
			\label{fig:boundingboxes}
		\end{figure}
	
		\begin{figure}
			\centering
			\includegraphics[width=\columnwidth]{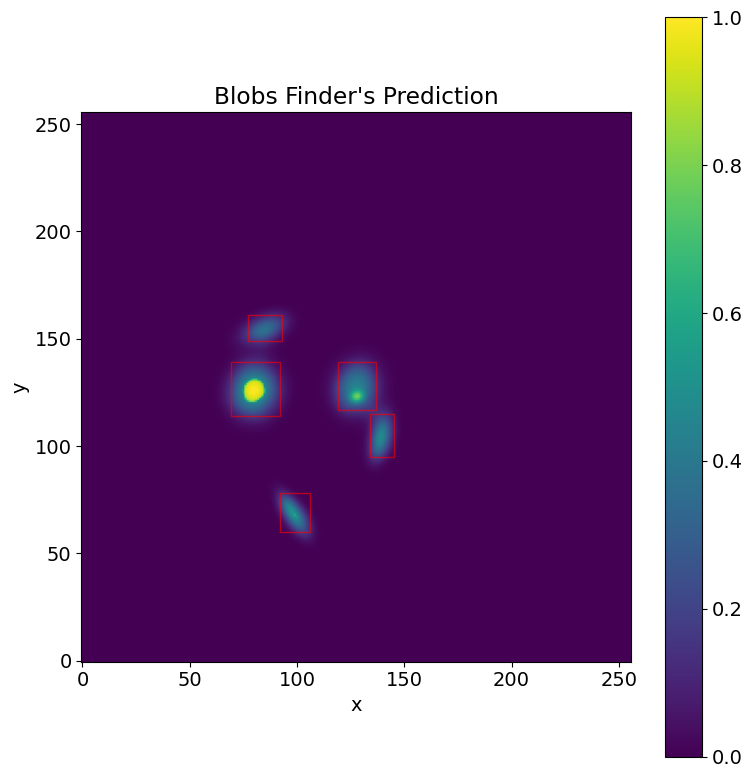}
			\caption{An example of Blobs Finder's output 2D probabilistic
				source detection map with the predicted bounding boxes
				extracted through thresholding, highlighted in red.}
			\label{fig:target_map}	
		\end{figure}

	\item \textbf{Frequency Denoising and Line Detection} (\textbf{5 - 7}): bounding boxes around source candidates are used to extract \textbf{dirty spectra} from the input cube. The spectrum for each source candidate is extracted by adding the pixels inside its bounding box over all the $128$ frequency slices of the cube. The spectra of the detected source candidates are extracted from the cube, standardized, i.e. rescaled to have null mean and standard deviation with unity value, and then fed to \textbf{Deep GRU} (Sec.~\ref{subsec:gru}). The Deep Gated Recurrent Unit denoises the standardized spectra and outputs 1D probabilistic maps of source emission lines or \textbf{cleaned spectra} (6).
	The cleaned spectra are then analysed with the \textit{scipy} \citep{scipy} \textit{find\_peaks} functions with a threshold value of $0.1$ in order to detect emission peaks. Each peak is then fitted with a 1D Gaussian model through the \textit{astropy} \citep{astropy} \textit{models.Gaussian1D} function.
	As fitting algorithm, we employ the \textit{LevMarLSQFitter} algorithm, and
	      as the initial values for the mean and amplitude of the 1D Gaussian model,
	      we use, respectively, the previously detected peak position and its relative
	      amplitude.
	      \begin{figure}
		      \centering
		      \includegraphics[width=\columnwidth]{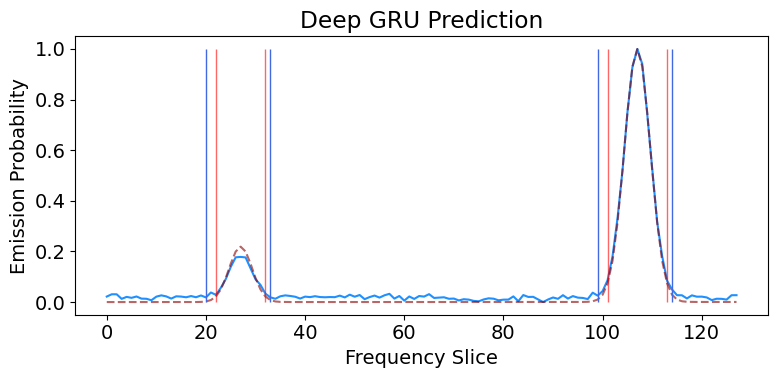}
		      \caption{In blue the dirty spectrum extracted from the central source bounding box predicted by Blobs Finder (Fig.~\ref{fig:boundingboxes}), in dotted-red the Deep GRU's prediction. Vertical blue bars delimit the true emission ranges, while red bars the predicted emission ranges. A secondary fainter source emission peak is detected by Deep GRU and thus the source is flagged for deblending.}
		      \label{fig:deep_gru_prediction}
	      \end{figure}
	    All detected Gaussian peaks position over the frequency axis ($z$) are recorded alongside their extensions $\Delta_z$ defined as $\Delta_z = 2* FWHM_z$ where $FWHM_z$ is the FWHM of the Gaussian peak (7).
	    In order to detect possible false positives produced by Blobs Finder, all
	      potential candidates showing no meaningful peak in their spectra are removed.
	      If more than one peak is found alongside the spectrum, the candidate is likely
	      the superimposition of blended sources and thus it is flagged for deblending.
	      All the remaining detections are then passed to the next stage. The peaks $\Delta_z$ are used to cut out the emission regions from  
	      the dirty spectra, which are fitted with a simple linear model implemented through the \textit{astropy models.Linear1D} function in order to take into account any possible trend of the continuum with frequency.
	      Fig.~\ref{fig:deep_gru_prediction} shows the dirty spectrum extracted from the two blended sources
		  showcased in Fig.~\ref{fig:target_map}, and the Deep GRU's predicted emission probability map. 
		  Blue vertical bars limit the true emission ranges of the two sources within the spectrum, while red ones limit the predicted emission ranges.

	\item \textbf{Source Spectral Focusing} (\textbf{8 - 9}): this phase has three main objectives: (i) to feed
	      to the ResNets the best possible input image of the potential candidates in order to
	      ease as much as possible the morphological parameters detection; (ii) to deblend the sources,
	      and (iii) to remove false detections. Regarding the first objective, the idea is to
	      increase as much as possible the Signal to Noise Ratio (SNR) of the source in the
	      image by cropping a $[64, 64]$ pixel box around its bounding box and integrate
	      it only within its peak FWHM. This operation is what we call \textit{spectral focusing} and the resulting image is the \textbf{spectrally focused image}.
	      In order to estimate the SNR of a source, we introduce two SNR measurements:
	      \begin{itemize}
		      \item \textbf{Global SNR}: we define the Global SNR as:\\
		            \begin{equation}\label{eq:global_SNR}
			            SNR = \frac{median(x_s(r))}{var(x_n(R - r))}
		            \end{equation}
		            where $x_s(r)$ are the values of the source pixels contained within
		            the circumference of radius $r$ that inscribes the source bounding box,
		            and $x_n(R - r)$ are the pixel values within an annulus of internal
		            radius $r$ and external radius $R$ which has the same area
		            of the inscribed circumference;
		      \item \textbf{Pixel SNR}: we define the Pixel SNR as:
		            \begin{equation}
			            snr = \frac{x_i}{var(X)}
		            \end{equation}
		            where $x_i$ is the value of the given pixel, and $var(X)$ is
		            the variance computed on the full image.
	      \end{itemize}
	      The two SNR measurements are used respectively to distinguish false detections from true sources, and to deblend overlapping sources within a blob. Fig.~\ref{fig:deblending_and_FP} summarises the false positive detection pipeline which works in the following way:
	      if a potential source has a Global SNR higher than $6$ (empirical bright source SNR threshold) in the integrated dirty cube, and was not flagged for deblending (due to the presence of multiple detected peaks in the extracted spectrum),
	      it is first focused and then passed along to the next stage of the pipeline; if it has a
	      SNR lower than $6$ in the integrated dirty cube, but there is an increase in the Global SNR when focused, it is also passed along to the next stage, otherwise,
	      it is discarded as a false detection (condition marked with $1$ in Fig. \ref{fig:deblending_and_FP}). The latter case, in fact, could only happen if the source is integrated outside its true emission peak, for example over a noise spike. 
       If more than one peak is found in the
	      potential source spectrum, then there is a chance that multiple blended sources make the blob (detected by BlobsFinder). 
	      First, the source is focused on the highest peak (primary peak) by integrating within its extension and the Global SNR calculation is made to understand if the potential source must be discarded (see previous step). Also, the Pixel SNR
	      measurement is used to find the highest SNR pixel in the image $p(x, y)$, which will act as a reference for the
	      next phase of the deblending process. 
        The potential source is then focused around the secondary peaks.
	      For each of these, if the peak is not superimposed
	      in frequency with the brightest peak, then by integrating within the extension of the secondary peak, the brightest source should disappear
	      from the focused image (because it is integrated outside its emission range), and the pixel
	      SNR measurement is used to find the highest SNR pixel in the image $s(x,y)$. If this pixel is different
	      from the previously found reference pixel (their distance is higher than a used defined number of pixels, in our case $3$), then the neighboring pixels
	      around this pixel are linked with a friend of friends algorithm in an iterative
	      manner. For each iteration, the Global SNR is measured.
	      Pixels are added in this fashion until a plateau or saturation level in the Global
	      SNR is reached. A bounding box is then created in order to encompass all the selected pixels, and
	      a $[64, 64]$ pixel image is cropped around the bounding box.
	      If, instead, the secondary peak overlaps the primary peak in frequency, and the two pixels coincide spatially ($p(x, y) = s(x, y)$), then the secondary peak is discarded as a false detection (i.e. not considered an independent source with respect to the primary, a condition marked as 2 in Fig. \ref{fig:deblending_and_FP}). If an increase in SNR is recorded by integrating over the multiple peaks, then the secondary peak is deemed as part of the primary source and the source emission range is extended accordingly, otherwise it is discarded as a noise spike (false detection). The first conditions may happen if DeepGRU mistakenly predicts a single true emission peak as two separate peaks or if Blobs Finder predicts a single true source as two very close blobs, while the latter if DeepGRU overpredicts the true emission range.
	   \begin{figure}
		      \centering
		      \includegraphics[width=\columnwidth]{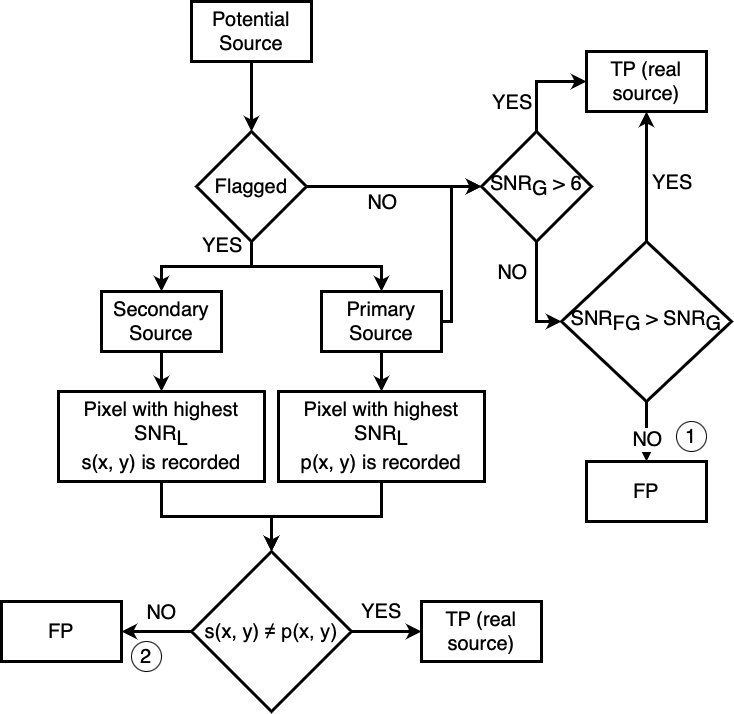}
		      \caption{Schema of the False Positives detection and source deblending pipeline which constitutes step $9$ of the source detection and characterisation pipeline (Fig.~\ref{fig:pipeline}). With the numbers $1$ and $2$ are marked the two possible conditions which may led to a potential source being defined as a false positive detection and thus discarded from further analysis. The under-script $FG$ (focused global) indicates that the global SNR is measured on the focused source, while $L$ implies a (local) Pixel SNR measurement. With \textit{flagged} we indicate that multiple peaks are detected within the potential source's spectrum and thus is flagged for deblending. For an in-detail description of this schema see Sec.~\ref{subsec:pipeline} (iii) Source Spectral Focusing.}
		      \label{fig:deblending_and_FP}
	      \end{figure}
	   Finally all spectrally focused sourced with a global SNR lower than $1$ are flagged and removed (9).
	   
       The deblending method assumes that all sources can be fairly approximated as 2D Gaussians (in the spatial plane) with
	      a single emission peak in frequency (1D Gaussian), and thus it will need be modified whenever our galaxy models will also simulate the velocity
	      dispersion and inclination along the line of sight. The introduction
	      of these parameters in our simulations would, in fact, create more complex spectral profiles,
	      for example a source with a high dispersion or inclination along the line of sight could show
	      multiple peaks in the spectrum. In this case, the simple logic that we have described above will need to be revised.
	      \begin{figure*}
		      \centering
		      \includegraphics[width=\textwidth]{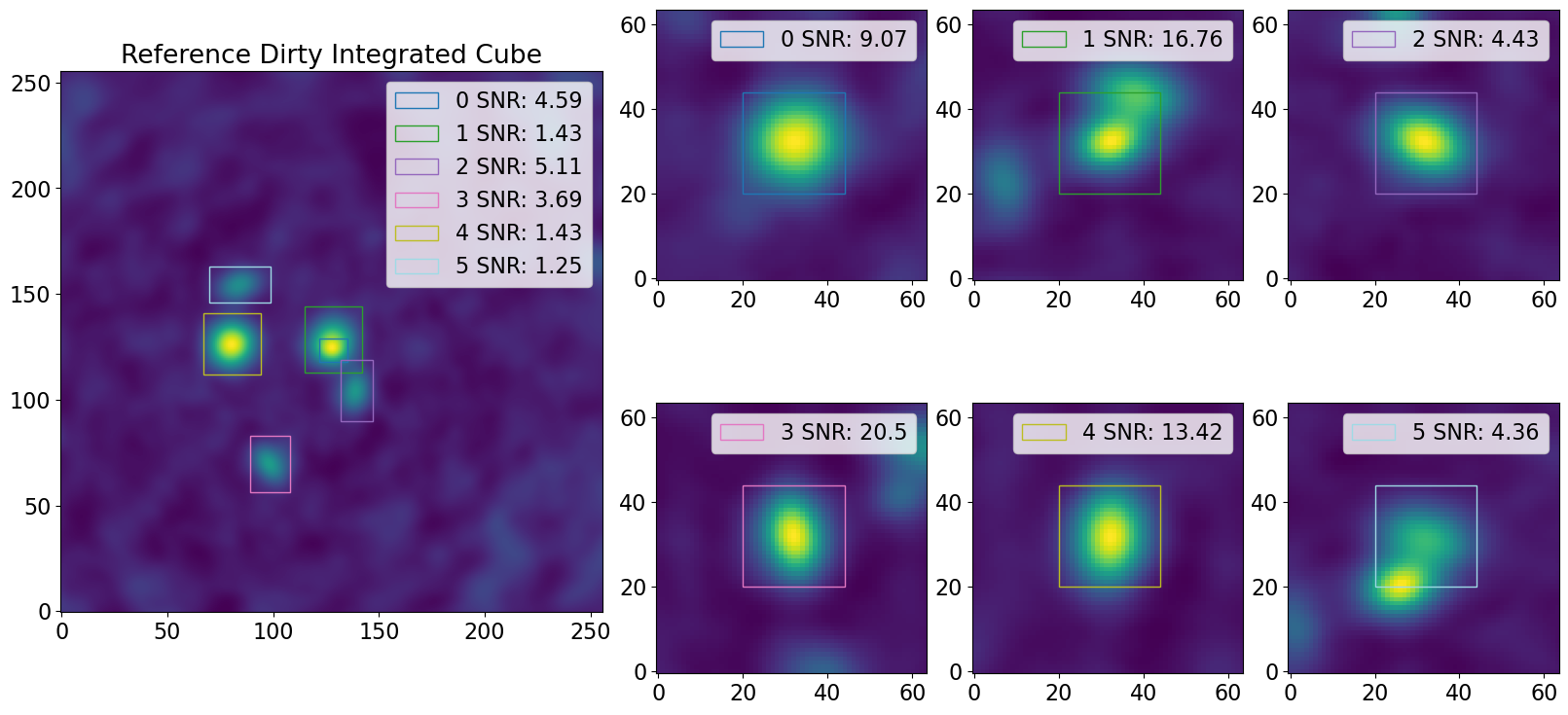}
		      \caption{An example of source spectral focusing of sources within a test set image. 
		      On the Left, as reference, we plot the dirty integrated cube with the
		      predicted 2D bounding boxes obtained by Blobs Finder highlighted in different colours. 
		      The legend matches the source number to the bounding box colour in the image and the measured Global SNR (see Eq.~\ref{eq:global_SNR}). 
		      On the right, there are the 6 Spectrally Focused images obtained by integrating over the predicted line extensions found by the Deep GRU and cropping a $[64, 64]$ pixel image around Blobs Finder's predicted bounding boxes centres. In each focused image it is also
		      showcased the measured Global SNR. As it can be seen there is a substantial increase in SNR when sources are focused around
		      their actual emission ranges.}
		      \label{fig:source_focusing_single_peak}
	      \end{figure*}

Fig~\ref{fig:source_focusing_single_peak} showcases the results of the Spectral Focusing of the potential sources detected
by Blobs Finder and Deep GRU in the test cube already displayed in Fig~\ref{fig:dirty_map}. 
By focusing on the two peaks detected by the Deep GRU (Fig~\ref{fig:deep_gru_prediction}, the two blended sources produce two different images (Focused Source 0 and 1) which now can be analysed independently. 
It is also worth noticing that by focusing the source within its effective emission range, signal from other sources is suppressed and noise variation is minimised resulting in a higher SNR than that registered in the reference dirty integrated image.
	      
\item \textbf{Morphological Parameters Estimation} (\textbf{10-11}): the spectrally focused images are normalized to the $[0, 1]$ range and then fed to three specialized ResNets (Sec.~\ref{subsec:resnet}). Each ResNet is specialized in predicting a given morphological parameter of the source in the image. Namely: the FWHM in the $x$ and $y$ directions and the projection angle ($pa$). The first two parameters are predicted in pixel values, while the angle is directly predicted in degrees (10) measured in a clockwise fashion.
The $x$ and $y$ positions of the source are computed as the photometric baricenters (pixel-weighted centers) of the Blobs Finder predicted bounding boxes. 
The predicted parameters are then used to generate a 2D model of the source, which is a 2D Gaussian generated at the source position with the predicted morphological parameters;
	\item \textbf{3D Model construction} (\textbf{12}): having fitted the source both in frequency and in the image plane, we combine the two Gaussians to create a 3D profile of the source, which then is converted into a 3D segmentation map, i.e. a binary cube with the same shape as the input dirty cube with all pixels that belong to the source set to $1$ and the rest set to $0$. To account for the fact that the
	      convolution with the dirty beam spreads both the continuum and the line emission
	      in the image, we dilate the segmentation map by a factor of $1.5$. This is
	      performed to make sure that all the source signal is contained within the 3D
	      segmentation map.
	\item \textbf{Flux Estimation} (\textbf{13}): the dilated 3D segmentation mask is then 
	      used to create the model-masked cube by multiplying it with the dirty cube.
	      The model-masked cube is, therefore, a version of the dirty cube in which all the pixels
	      outside the source 3D model are set to zero.
	      The inverse mask is instead used to capture the continuum cube of shape
	      $[128 - \Delta_z, 256, 256]$ where $\Delta_z$ is the size of the segmentation
	      mask in frequency channels. The continuum image is then created by averaging the continuum cube in frequency and the line emission cube is created through the following formula:
	      \begin{equation}
		      L_z[x, y] = M_z[x, y] - f(z) * C[x, y]  \quad \textrm{with} \  z \in \Delta_z
	      \end{equation}
	      where $L_z[x, y]$ it the 2D line emission image at slice $z$, $M_z[x, y]$ is the
	      model masked 2D image at slice $z$, $C[x, y]$ is the continuum image and
	      $f(z)$ is the 1D continuum model. The line emission cube is integrated along the frequency to create the line emission
	      image which is fed to a specialized ResNet predicting the source
	      flux density in mJy/beam.
\end{enumerate}

\subsection {Training Strategies}\label{subsec:training}
In the pipeline, the data flows from one DL model to the next. In particular, the dirty spectra extracted from Blobs Finder predicted model images are the inputs of Deep GRU; the
outputs of Deep GRU - the denoised spectra - are  combined to produce the \textit{focused}
images used as input for the parametric ResNets, and the 3D models constructed with
the ResNets parameters prediction is used to produce the line images, which are the input of the
flux ResNet. Since we work with simulated data, both the true model images and the parameters of sources within them are known to us and thus all models within the pipeline can be trained at the same time on different GPUs by preemptively using the true source parameters to extract from the cubes the needed spectra and spectrally focused images, and the line emission images to train respectively Deep GRU and the ResNets. The problem with this training strategy is that it would not take into account the fact that Deep GRU and the ResNets, in production, will not receive perfectly extracted spectra or focused images but the product of the imperfect prediction of the previous models in the pipeline schema. Deep GRU could receive (as input) spectra which contain only partially the source emission, and the ResNets not perfectly focused images. To account for this, we first trained all models in parallel in order for them to learn how to solve their respective problems while optimising the pipeline total training time (which benefits from the fact that the models training are carried on at the same time), and then we also trained each model (with the same training strategies that we will outline for each model, and with the exclusion of Blobs Finder which is the first model in the pipeline) on the un-augmented training set predictions of the previous model. In this way each model should be able to correct for the mistakes (biases in the data) of the previous one.  

Blobs Finder is trained on pairs of dirty input images (input dirty data cubes integrated along frequency) and target sky model images (target sky model cubes integrated along frequency).
To achieve translational and rotational invariance in the network, at each iteration, both input and target images are randomly rotated by an angle $\theta$ with $\theta \in [0^{\circ}, 360^{\circ}]$,
randomly flipped in the $x$ and $y$ planes with a probability of $1$ and then cropped from $[360, 360]$ pixels to $[256, 356]$. The cropping operation brings no loss of data, given that all the values in images outside the primary beam of the simulated ALMA observations ($r \sim 138$ pix) are set to \textit{nan} values, and those at the edges of the beam have unreliable SNR.
After cropping, both integrated dirty cubes and target sky model images are normalized to the $[0, 1]$ range before being fed to Blobs Finder.

The model is trained with an Adam Optimiser \citep{adam}, which is the optimization algorithm used to update the weights of the model
on the basis of the loss function. The amount of change imparted to the model's weights to minimize the loss is moderated through the \textit{learning rate}. Adam utilises a different learning rate for each parameter and updates them on the basis of the first and second moments of the gradients. Another problem in DL model's training, is the possibility of overfitting the data batches in the first training iterations, given the model's inherent initial
instability due to the random initialisation of its weights. 
To prevent that,
we adopt a \textit{warm up} strategy for the learning rate \citep{warm-up} in which we start with a learning rate of $0$, and we uniformly increase it to  $1 \times 10^{-4}$ in the first $10$ iterations.
The model is then trained for a maximum of $300$ epochs, but we also employ an early stopping criterion based on the validation loss.
If no improvement of validation loss with respect to the moving average of the last $10$ validation losses
is registered for $10$ consecutive steps, then training is halted. As outlined in Sec.~\ref{subsec:blobsfinder},
Blobs Finder is trained with the weighted combination of the $l_1$ loss and the $DSSIM$ loss.
\begin{equation}
	l(x, y, t) = a(t) \times l_1(x, y) + b(t) \times DSSIM(x, y)
\end{equation}
where $x$ and $y$ are the prediction and target variables and $t$ is the epoch counter.
The weighting is performed with two variables $a$ and $b$, which depend on the epoch counter $t$.
The training begins with $a(0) = 1$ and $b(0) = 0$ in order to first allow the model to learn a \textit{median} representation of the data (which should contain information about the PSF and noise patterns,
assumed roughly constant in the data) through the minimization of the $l_1$ loss. At each epoch, $a$ is decreased by $\delta$ and $b$ is increased by the amount in order to slowly transition to the $DSSIM$ loss. In the last epochs, only the $DSSIM$ loss is effectively used in order to to learn the nuances in the data, such as source positions or morphological properties (the shape and sizes of the galaxies). 

Deep GRU is trained on pairs of dirty spectra (extracted from the dirty cubes) and clean spectra (extracted from the sky model cubes) through the bounding boxes obtained from Blobs Finder Prediction (or the known parameters in case of parallel training)
The input dirty spectra are standardized while the target model spectra are normalized to the $[0, 1]$ range.
To train the model, we again use the Adam optimization algorithm, a warm-up strategy for the learning rate with
a working learning rate of $1 \times 10^{-4}$ and a weight decay of $1 \times 10^{-5}$ to prevent the model from overfitting the training set. The model is trained for $300$ epochs on spectra extracted through the known parameters  and for $50$ iterations on spectra extracted from Blobs Finder's predictions. We also employ an early-stopping criterium on the basis of the validation loss. 
The Deep GRU predictions are used
in combination with Blobs Finder's predictions to extract the \textit{spectrally focused} galaxy images,
i.e., $64 \times 64$ pixels images with the source roughly in the centre and normalized to the $[0, 1]$ range (see Sec.~\ref{subsec:pipeline} for more details).
As targets for training, we use the true source parameters used to produce our simulations.
The three ResNets for morphological parameters estimation are trained simultaneously for $300$ iterations with the same stopping criterion set for Blobs Finder's training on perfectly spectrally focused images (obtained through the know source parameters) and for $50$ iterations on the spectrally focused sources obtained though Blobs Finder and Deep GRU's predictions.
Finally, the outputs of the ResNets (i.e., the sources' morphological parameters) are used to construct 3D models of the galaxies, which are used to create the segmentation masks from which we measure
the continuum and create the line emission cubes and then the line emission images (for the details, see Sec.~\ref{subsec:pipeline} points \textit{v} and \textit{vi}). 
The last ResNet is trained with the line emission images as inputs and the fluxes as targets. We use the same training strategy outlined for the other
ResNets.

\section{Source Detection and Characterisation}\label{sec:source_detection_and_characterization}
In this section, we present the performances of the different steps of the pipeline over $1,000$ cubes which belong to the Test set. Whereas possible we also compare the performances of our pipeline with those of other pipelines widely used in the community:  \textsc{blobcat} \citep{blobcat}, \textsc{Sofia-2} \citep{sofia} and \textsc{decoras} \citep{Rezaei2021}.

\subsection{Source Detection}\label{sec:source_detection}
We start with Blobs Finder's performance in detecting sources within the 2D Test set integrated dirty cubes.
Following Sec.~\ref{subsec:pipeline} step $1$, the output 2D Probabilistic Maps are binarized through a hard
threshold of $0.15$, and bounding boxes around islands of connected pixels are extracted.
To check if a source has been detected by Blobs Finder, we measure the 2D Intersection over Union (2D IoU) between the true 2D bounding box and the predicted one, while for Deep GRU the 1D IoU is measured between the true emission ranges and the detected ones. In both cases, a threshold of $0.6$ is used. To ensure that the central part of the source emission of a True Positive (TP) is always detected, we require the distance between the centres of the true and predicted bounding boxes to be smaller then $3$ pixels. The combination of these two thresholds should ensures that at least $60\%$ of the 3D emission range of a source must be captured in order for its prediction to be deemed a TP. Also, given that the line emission image is created through the dilated segmentation mask, the $0.6$ IoU threshold guarantees that $90\%$ of the true emission range is captured within it.  

\begin{figure*}
	\centering
	\begin{subfigure}[b]{1\textwidth}
		\includegraphics[width=\textwidth]{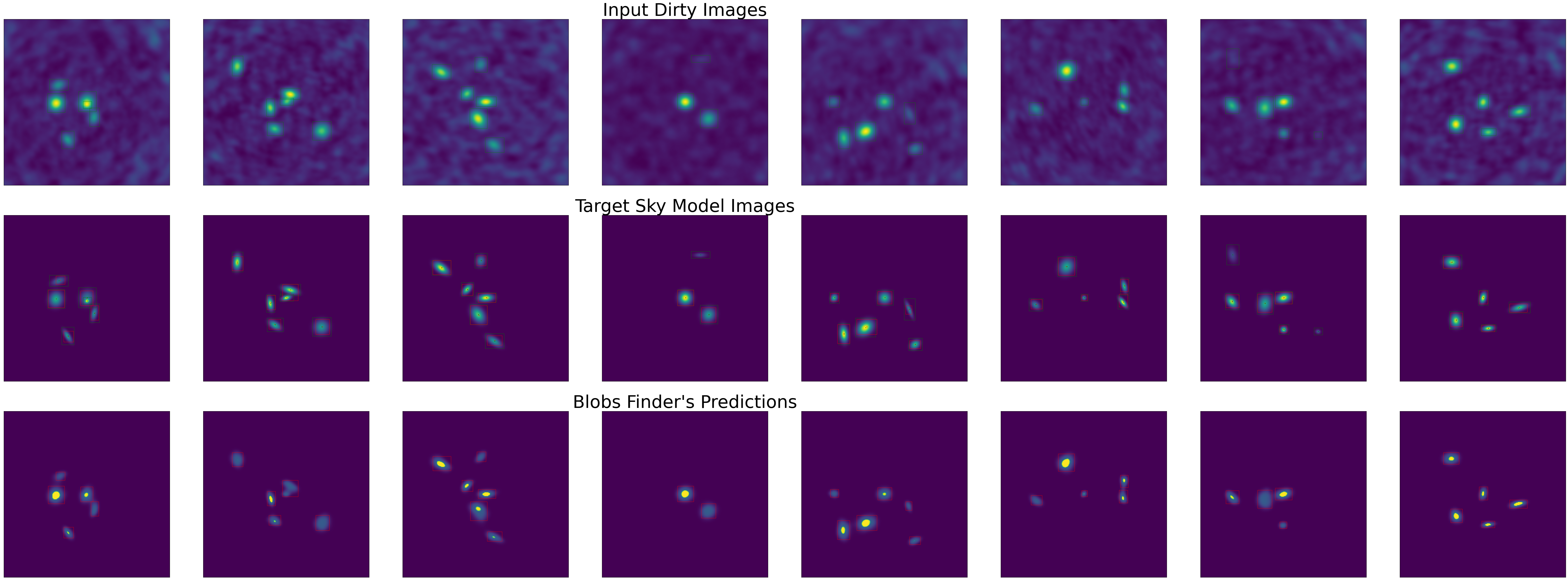}
		\subcaption{}
	\end{subfigure}
	\begin{subfigure}[b]{1\textwidth}
		\includegraphics[width=\textwidth]{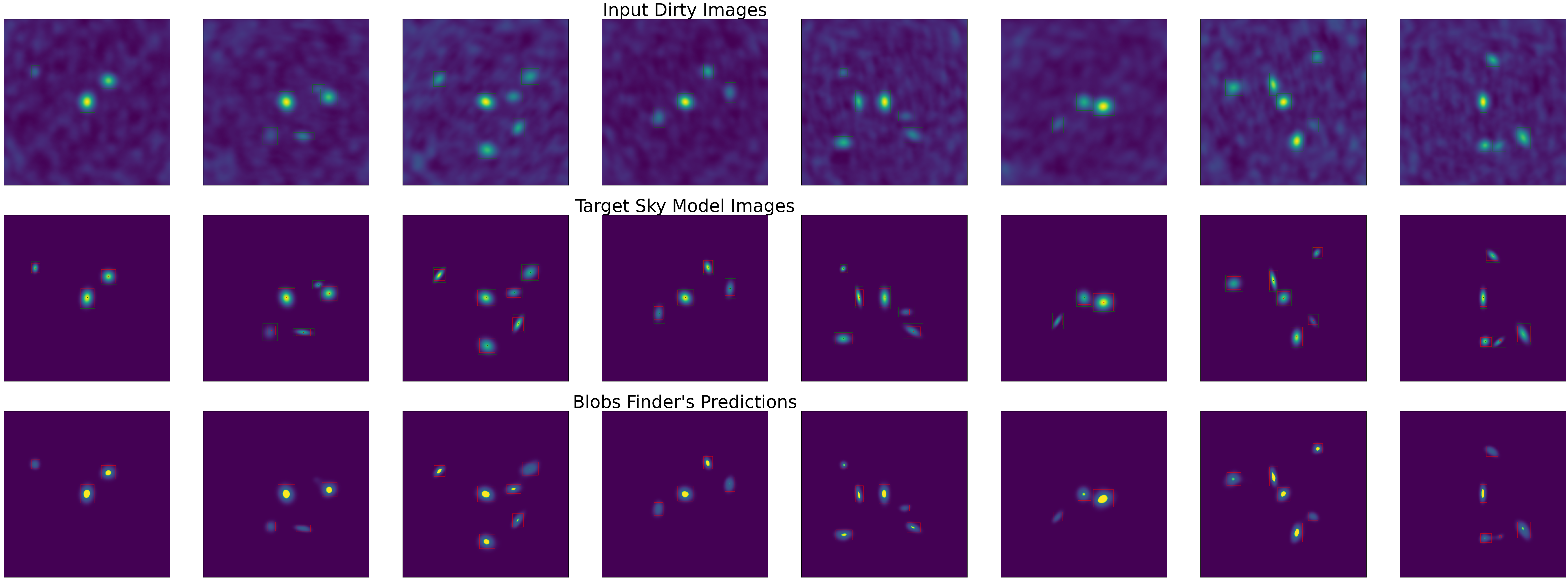}
		\subcaption{}
	\end{subfigure}
	\caption{Examples of Blobs Finder predictions on the test Set. The first row shows input integrated dirty cubes, the middle row the target sky models, and the bottom row, Blobs Finder predicted 2D Source Probability maps. In green are outlined (in the dirty and sky models images) the true bounding boxes, while in red the predicted bounding boxes extracted by thresholding the probability maps.}
	\label{fig:blobsfinder_examples}
\end{figure*}

\begin{figure*}
	\centering
	\begin{subfigure}[b]{1\textwidth}
		\includegraphics[width=\textwidth]{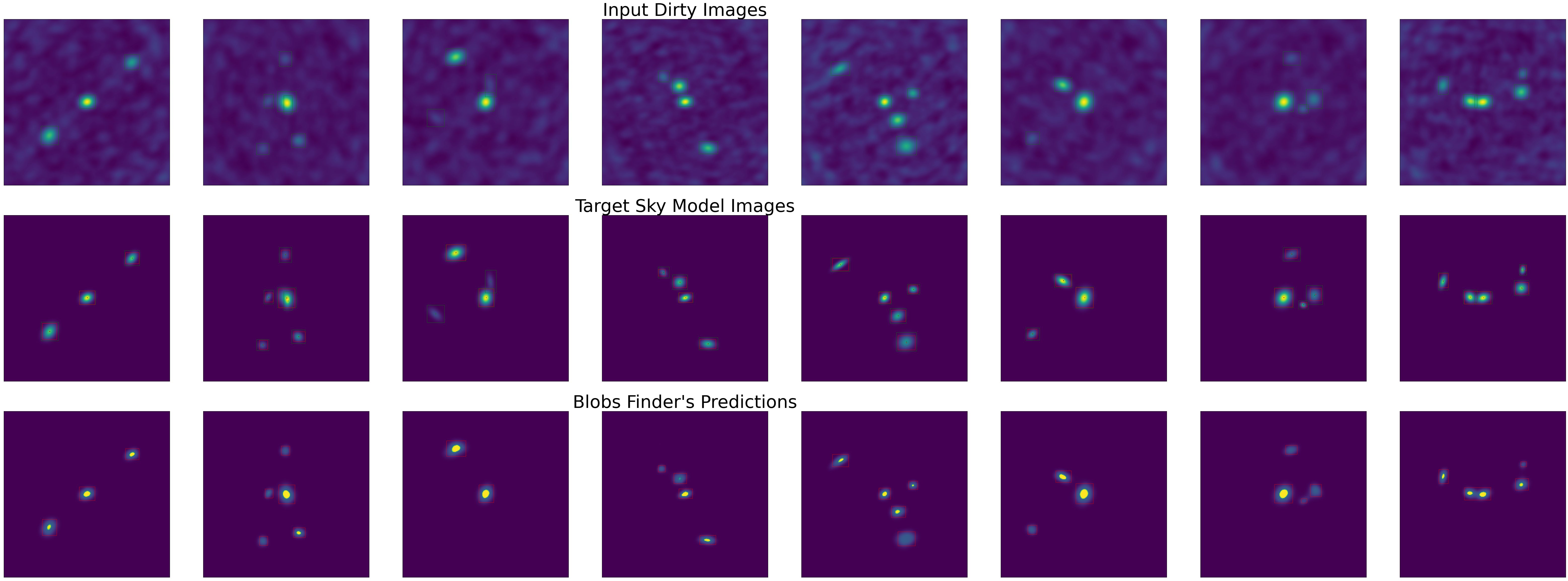}
		\subcaption{}
	\end{subfigure}
	\begin{subfigure}[b]{1\textwidth}
		\includegraphics[width=\textwidth]{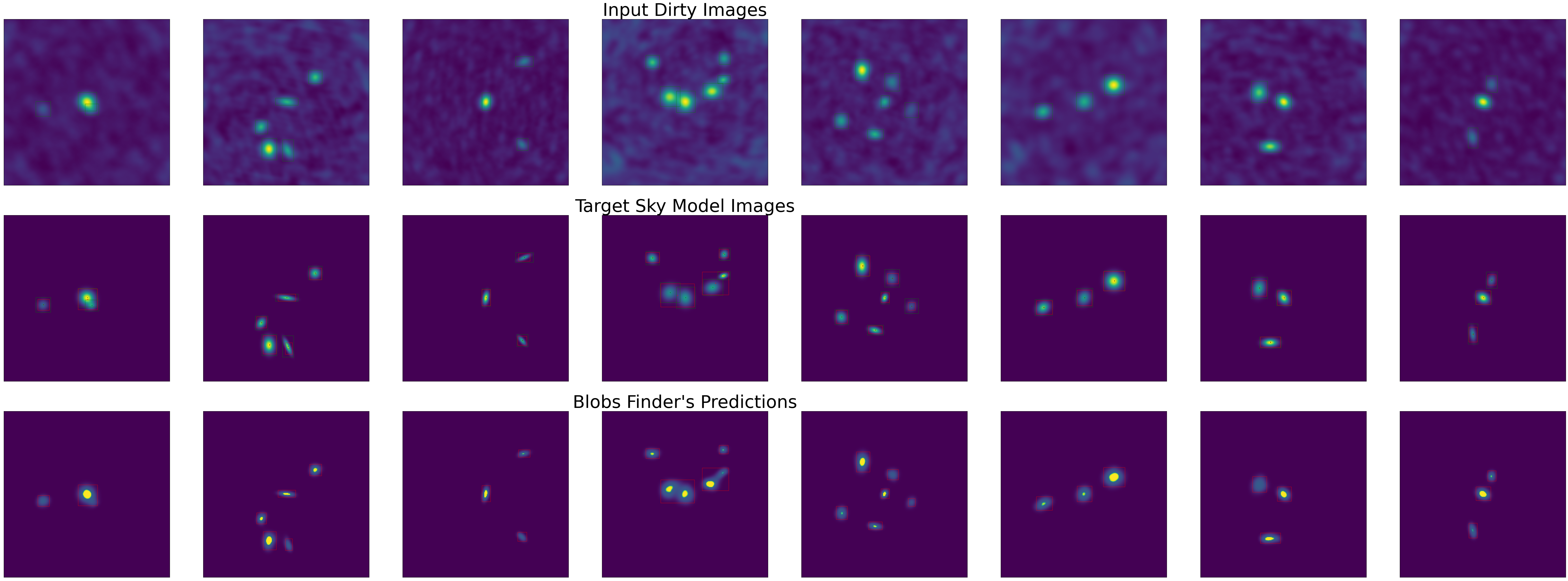}
		\subcaption{}
	\end{subfigure}
	\caption{Examples of Blobs Finder predictions on the Test Set. The first row shows input integrated dirty cubes, the middle row the target sky models, and the bottom row, Blobs Finder predicted 2D Source Probability maps. In green are outlined (in the dirty and sky models images) the true bounding boxes, while in red the predicted bounding boxes extracted by thresholding the probability maps.}
	\label{fig:blobsfinder_examples_1}
\end{figure*}

Getting a good estimate of the centers is important given that, except for blended sources that potentially will get a new center later on in the pipeline based on the SNR reasoning outlined in Sec.~\ref{subsec:pipeline} step $3$, the ResNets will receive as input focused images cropped around the  bounding boxes centers predicted by Blobs Finder. Hence, the closer the centers are to the true centers, the more sources will be centered in the focused images. 
As an example, by looking at the two blended sources in Fig.~\ref{fig:boundingboxes}, one can see that the predicted bounding-box (in red) encompasses the most of the emission of the extended faint source and the totality of the emission of the compact
source, so there is the potential of detecting both sources in the spectrum. When applied to the test set,
Blobs Finder predicts $4056$ ($89\%$) sources (TP) against the true $4,556$ sources (using both the distance and IoU criteria). $4,205$ ($92.3\%$) sources pass the 2D IoU criterion only, meaning that an additional $149$ sources are detected by Blobs Finder but are spatially blended with another source. Blobs Finder also misses $354$ sources (FN) and detects $4$ False Positives (FP). 
The $4056$ bounding boxes are used to extract a corresponding number of dirty spectra from the dirty cubes. The Deep GRU detects $4,202$ emission peaks out of the $4,205$ present in the extracted spectra but also produces $62$ false positives. To detect and remove false detections and confirm true ones, sources are ''spectrally focused'' within the predicted frequency emission ranges $\Delta_z$, and SNR checks are made (see Sec. \ref{subsec:pipeline} (iii) Source Spectral Focusing). The full logic of the FP removal process is shown in Fig.~\ref{fig:deblending_and_FP}. Regarding the $4$ FPs detected by Blobs Finder, when the spectra extracted from the respective bounding boxes are feed to DeepGRU, it detects no emission peak within $3$ of the $4$, and thus the first $3$ FPs are discarded as false detections. The last FP is eliminated through condition $2$ of the schema shown in Fig.~\ref{fig:deblending_and_FP}. Regarding the $62$ FP peaks detected by DeepGRU, $29$ were artificial peaks predicted near the spectral borders (the boundaries of the spectra may be recognised as peaks given the presence of a discontinuity in the signal) and, after being identified as primary sources (given the higher amplitude with respect to the real peaks), they failed the global SNR test when focused in their emission range (condition $1$ in Fig.~\ref{fig:deblending_and_FP}) and thus were discarded as false detections. Of the remaining $33$ false detections made by DeepGRU,  $31$ are eliminated through condition $2$ Fig.~\ref{fig:deblending_and_FP}. These false peaks are all superimposed with their respective primary peaks in the spectra and the pixels with the highest SNR in the primary and secondary peak's spectrally focused images are less than $3$ pixels apart and thus were considered either as noise spike or as part of the primary source. The remaining $2$ false peaks produced specrtally focused images with a measured a global SNR less then $1$ and thus are flagged as false detections.
Fig.~\ref{fig:blobsfinder_examples} and \ref{fig:blobsfinder_examples_1} show some examples of Blobs Finder predictions on the test set. For each block, the first row shows the input integrated dirty cubes, the middle row the target sky models, and the bottom row, Blobs Finder predictions.

To compare with \textsc{blobcat} (\cite{blobcat}) and \textsc{Sofia-2} (\cite{sofia}), we run both algorithms on the $1000$ dirty images in the test set. Blobcat requires two parameters: a detection ($T_d$) and cut ($T_f$) SNR threshold to decide which peaks in the image are good candidates for blobs and where to cut the blobs boundaries around them (in other words: pixels with a SNR higher than $T_f$ are selected to form islands and island boundaries are defined by $T_d$). To make the fairest comparison possible, we measured \textsc{blobcat} performances with different choices of $T_d$ and $T_f$ through a grid-search strategy ($T_d \in [2, 15]$, $T_f \in [1, 10]$) and, in this work, we report the best obtained results ($T_d = 8\sigma$, $T_f = 4\sigma$). 

The same criterion used for Blobs Finder was used to measure  \textsc{blobcat} performances. \textsc{blobcat} successfully detects $2,779$ ($61\%$) sources, produces $2,429$ false detection and misses $1777$ sources. The majority of sources missed by \textsc{blobcat} are spatially blended with brighter sources, or present a $SNR <= 5.0$, or are located at the edges of the images.
The \textsc{Sofia-2} smooth and clip algorithm (S + C) algorithm works by iteratively smoothing the data cube on multiple spatial and spectral scales to extract statistically significant emission above a user-specified detection threshold on each scale. 

Guided by the spatial and frequency sizes of the simulated sources, we employed spatial and frequency kernel sizes of $[3, 5, 7, 9, 11]$, and a source finding threshold of $0.5$. The algorithm proceeds by linking together meaningful detections with a friend of friend algorithm; we employed a grid-searching strategy to find the best possible spatial linking radii in the image and frequency dimensions of the cube within the interval $[1, 5]$ pixels. Finally the algorithm removes false positives on the basis of a reliability score based on the source SNR. We employ a SNR threshold of $2$ and we let the algorithm automatically select the other thresholds. We limit the detection area of \textsc{Sofia-2} by setting the masking to a $256$ pixel size square centered in the image. This is performed in order to cut the low SNR outskirt of the dirty images. \textsc{Sofia-2} detects $1010$ ($22\%$) sources, produces $4011$ false detections and misses $3546$ sources. Most false positives are located at the spatial edges of the image, while it misses most blended sources by merging their emissions with other sources which leads to all involved sources failing the 2D IoU or distance based thresholds.

\begin{figure*}
    \centering
    \includegraphics[width=1\textwidth]{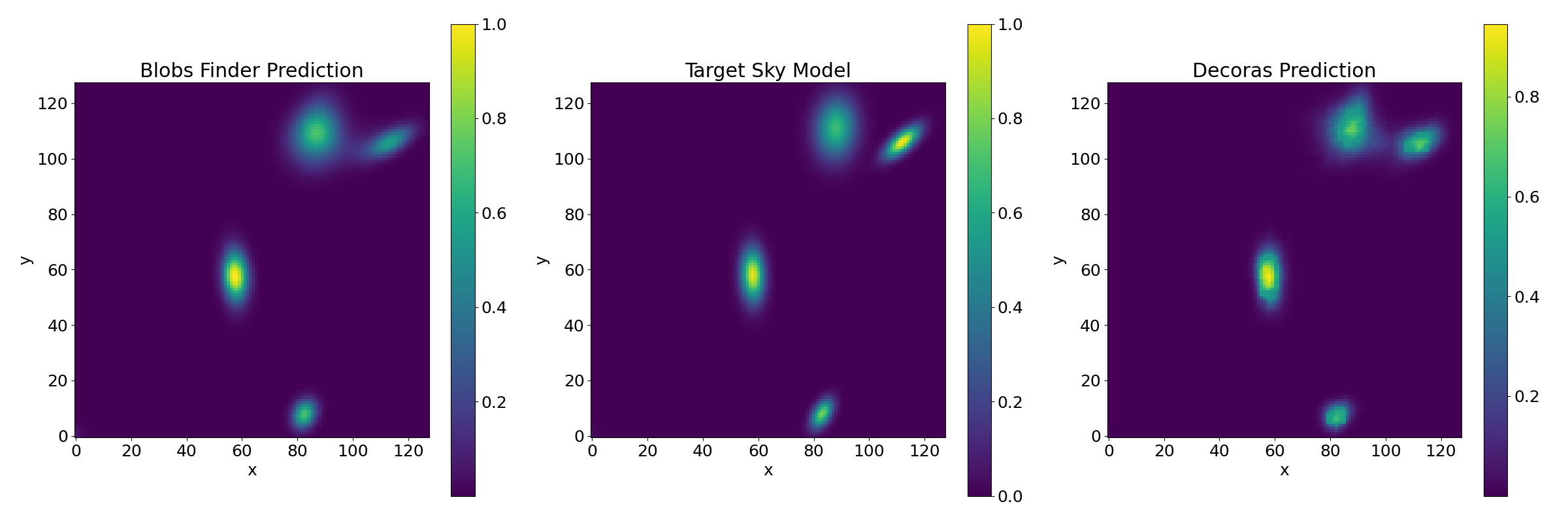}
    \caption{Left: Blobs Finder predicted 2D probabilistic map; Center: true sky model image; Right: \textsc{decoras} implementation of Blobs Finder predicted 2D probabilistic map. Predictions and target images have been cropped to $128$ by $128$ around sources, to better showcase the reconstruction quality.}
    \label{fig:reconstruction}
\end{figure*}

The \textsc{decoras} (\cite{Rezaei2021}) pipeline is constituted by two Deep Convolutional Autoencoders, the first one works exactly like our Blobs Finder, the second one takes $[128, 128]$ pixel cropped images around the first Autoencoder's predictions and predicts the source structure which is then fitted with a 2D Gaussian Function to find the source morphology. Our simulated cubes contain multiple sources while \textsc{decoras} assumes that there is a single source in each cube and thus the characterization part of the pipeline cannot be used to detect sources. For such reason, we compare our results with those by \textsc{decoras} only on  the detection part of the problem. First of all, following the guidelines outlined in their article, we re-implemented their Convolutional Autoencoder. The main architectural differences between our Convolutional Autoencoder and theirs are: their latent space contains $256$ atoms, while our $1024$, they perform spatial upsampling by only using chained Transposed Convolutions, while we use a combination of Bilinear Interpolation, Transposed Convolutions and Convolutions, and their Encoder and Decoder contain half our convolutional layers. In their paper, they train their Autoencoder with two loss functions: the Binary Cross Entropy (BCE) loss and the Mean Squared Logarithm loss (MSLE), and deem as correct predictions only those sources which are predicted by both models and whose distance (measured from the source centers found through the \textit{blob\_dog} \textit{scikit-image} algorithm) is less then $3$ pixels. 
We trained and tested their model with both loss functions using the same criteria outlined in Sec.~\ref{subsec:training} for Blobs Finder. After seeing that no meaningful result could be obtained with the BCE, we only use the model trained with the MSLE to perform the source detection on the $1000$ dirty integrated images in the test set. Given that the \textit{blob\_dog} algorithm only finds coordinates of the sources and does not output their emission boundaries, needed to create bounding boxes with which measure the IoUs, we make two performance measurement.
The first one relaxes the source detection criterion and only uses the distance-based threshold, while the second one uses both the distance and IoU criteria. To get bounding boxes from \textit{blob\_dog} outputs we use, for each blob detected by the algorithm, the best matching kernel standard deviations to create a radius of emission from which derive bounding boxes with the following equation:
\begin{equation}
    r = \sqrt{(3 * \sigma_b)^2}
\end{equation}
where $r$ is the obtained radius, and $\sigma_b$ is the standard deviation of the best fitting kernel. \textsc{decoras} detects $4,100$ ($89.9\%$) sources, produces $759$ false detections and misses $456$ ($10.1\%$) sources  using only the distance based threshold. 
The number of detected sources drops to $3,560$ 
($78.2\%$) if also the IoU-based threshold is used, while the number of false positives stays the same, and the number of missed sources rises to $996$ ($21.9\%$).  In Table~\ref{tab:detection_results} we summarise the source detection performances for all methods.
The two DL-based pipeline achieve improved performances over the traditional counterparts. By taking a look at \textsc{decoras} performances, especially if only the distance based criterium is used, one can see that they are similar to those of our Blobs Finder with respect to the number of TP while the number of FP is much higher. This behaviour could be connected to the use of subsequent Transposed Convolutions to perform spatial upsampling in the Decoder part of the network \citep{wojna} which may led to artefacts be mistaken as blobs by \textit{blob\_dog}.
Regarding the FP detection and removal, we show both the performances without the use of FP detection and removal pipeline (BF + DeepGRU), and those with it (Pipeline).\\ 
Blobs Finder and Deep GRU have a False Positive Rate (FPR) respectively of $\sim 10^{-3}$ (Blobs Finder) and $\sim 10^{-2}$ (Deep GRU) and a combined number of $66$ FPs which is already within acceptable limits. The low rate of FPs and the successful removal of all FPs by the DL pipeline are closely related to the nature of the mock data used for this study. While the false detections made by Blobs Finder were trivial to identify and remove, and may continue to be so in case of real data (no relevant peak was detected in the corresponding spectra by Deep GRU), the logic we used in removing the false detections made by Deep GRU that were due to a single true emission peak being detected as two superimposed peaks, will require further investigation.
In fact, if the velocity dispersion and inclination with respect to line of sight are taken into account (both factors are not yet present in our current simulations), the hypothesis that the spatial position of the brightest pixel of a source should remain constant within its spectral emission range no longer holds. Thus, the criterion which we employ to separate a true secondary peak from a false detection cannot be used anymore and needs to be modified.
Blobs Finders' and \textsc{decoras}' low numbers of FPs with respect to \textsc{Sofia-2} and \textsc{blobcat} are explainable due to the DL models capabilities of approximating the dirty beam. While Blobs Finder improved performances over \textsc{decoras} are due to architectural choices and training strategies, we believe that if dirty beam variations are increased by simulating multiple antenna configurations and observing conditions (integration time or azimuth), the number of FPs detected by both models will probably increase due to the increased complexity required to approximate a more realistic variating PSF.
\begin{figure*}
    \centering
    \includegraphics[width=1\textwidth]{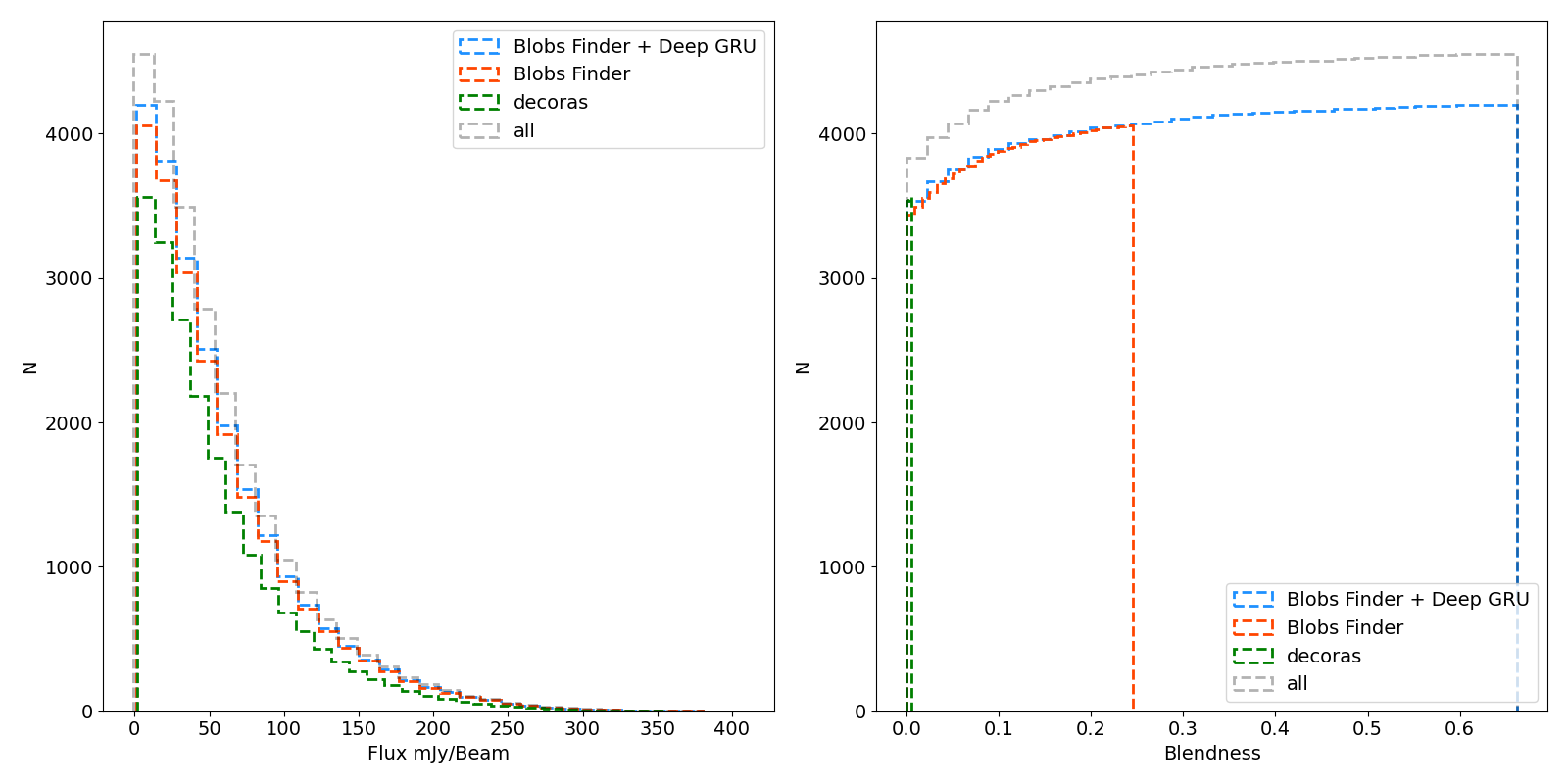}
    \caption{Left: histograms of the detected sources flux densities. Right: cumulative histogram of the detected sources \textit{blendness score} (see the text). In both histograms, we compare our detection pipeline (Blobs Finder + Deep GRU), our implementation of Blobs Finder, \textsc{decoras} implementation of Blobs Finder and,  we report the histograms for all the test set distribution.}
    \label{fig:pipeline_vs_decoras}
\end{figure*}
Fig.~\ref{fig:pipeline_vs_decoras} shows the histograms of flux densities (left) and  the \textit{blendnees scores} (right) of the detections made by our pipeline, Blobs Finder, and by \textsc{decoras}. The blendness score is defined as the maximum IoU between a source true bounding box and all other source bounding boxes within an image and can be used as a proxy to quantify how much different sources are spatially blended. Mathematically we define it as:
\begin{equation}
    b_i = max(IoU(box_i, box_j)) \; for \; j \in [1, N]
\end{equation}
where $N$ is the total number of sources within the image. 
If the blendness is $0$, it means that the source  is spatially isolated in the cube, while a blendness of $1$ means that the source is spatially superimposed to at least another source in the cube. 
We also report all the test set values for comparison. 
Our implementation of Blobs Finder is equivalent to \textsc{decoras}'s implementation regarding the minimum detected flux density of $1.31$ mJy/beam  but it seems to be more effective in deblending sources. Blobs Finder detects sources up to a blendness value of $0.232$, while decoras reaches at most $0.021$. It has to be said that also this difference in performance could be connected to the different ways the two pipelines extract bounding boxes from the model images produced by the Autoencoders and not to the quality of the images themselves.
To eliminate the effect of the extraction algorithm from our comparison, we compare the measured mean structural similarity index (mSSIM) between the true sky model test set images, Blobs Finder predictions and \textsc{decoras} predictions. Blobs Finder achieved a mSSIM on the test set of $0.003$, while our implementation of \textsc{decoras} achieved a mSSIM of $0.008$. Fig.~\ref{fig:reconstruction} shows the direct visual comparison between the true target sky model image (center), Blobs Finder (left) and \textsc{deoras} (right).

Deep GRU (and Source Spectral Focusing) does not improve on the minimum detected flux, but deblends sources and in fact pushes the maximum detected source blendness to $0.66$ which is the maximum blendness simulated in the data.

\begin{table}
    \centering
    \begin{tabular}{l|c|c|c }
        \hline
        \textbf{Algorithm} & \textbf{TP / } & \textbf{FP}  & \textbf{FN}\\
        \hline
        \textbf{BF + DeepGRU} & $4202$ ($92.3\%$) & $63$ & $354$ ($7.7\%$)\\
        \textbf{Pipeline} & $4202$ ($92.3\%$) & $0$ & $354$ ($7.7\%$)\\
        \textbf{blobcat} & $2779$ ($61\%$) & $2429$  & $1777$ ($39\%$)\\
        \textbf{Sofia-2} & $1010$ ($22\%$)& $4011$ & $3546$ ($78\%$)\\
        \textbf{decoras} & $3560$ ($78.2\%$)& $759$ & $996$ ($21.9\%$)\\ 
        \hline
        \textbf{Algorithm} &  \textbf{Precision} & \textbf{Recall} & \textbf{Mean IoU}\\
        \hline
        \textbf{BF + DeepGRU} & $0.98$ & $0.923$ & $0.74$\\
        \textbf{Pipeline} &  $1.0$ & $0.923$ & $0.74$\\
        \textbf{blobcat} & $0.53$ & $0.609$ & $0.61$\\
        \textbf{Sofia-2} & $0.20$ & $0.22$ & $0.63$\\ 
        \textbf{decoras} & $0.82$ & $0.78$ & $0.60$\\ 
        \hline
    \end{tabular}
    \caption{Comparison between the sequential application of  Blobs Finder and DeepGRU (BF + DeepGRU), the sequential pipeline completed with the Spectral Focusing for FPs removal and deblending (Pipeline), \textit{blobscat}, \textsc{Sofia-2} and \textsc{decoras}. Columns show true positives (TP), false positives (FP), false negatives (FN), precision, recall and mean intersection over union (Mean IoU) between true bounding boxes and predicted ones. TP and FN are also expressed as fractions over the total number of sources.}
    \label{tab:detection_results}
\end{table}


\subsection{Source Characterization}\label{sec:source_characterization}

The $4,202$ focused images are fed to the $3$ ResNets (steps $10$ and $11$ of the pipeline) each one fine tuned to predict one of the three morphological parameters:  $FWHM_x$, $FWHM_y$, and $pa$. The source positions $x$ and $y$ are computed as the pixel weighted centres of the sources bounding boxes, while the peaks frequency positions $z$ and extensions $\Delta_z$ are computed by fitting 1D Gaussians to the clean peaks found by Deep GRU. The sources parameters are used to create the Line Emission Image (see \ref{subsec:pipeline} Flux Estimation)

\begin{figure*}
    \centering
    \includegraphics[width=\textwidth]{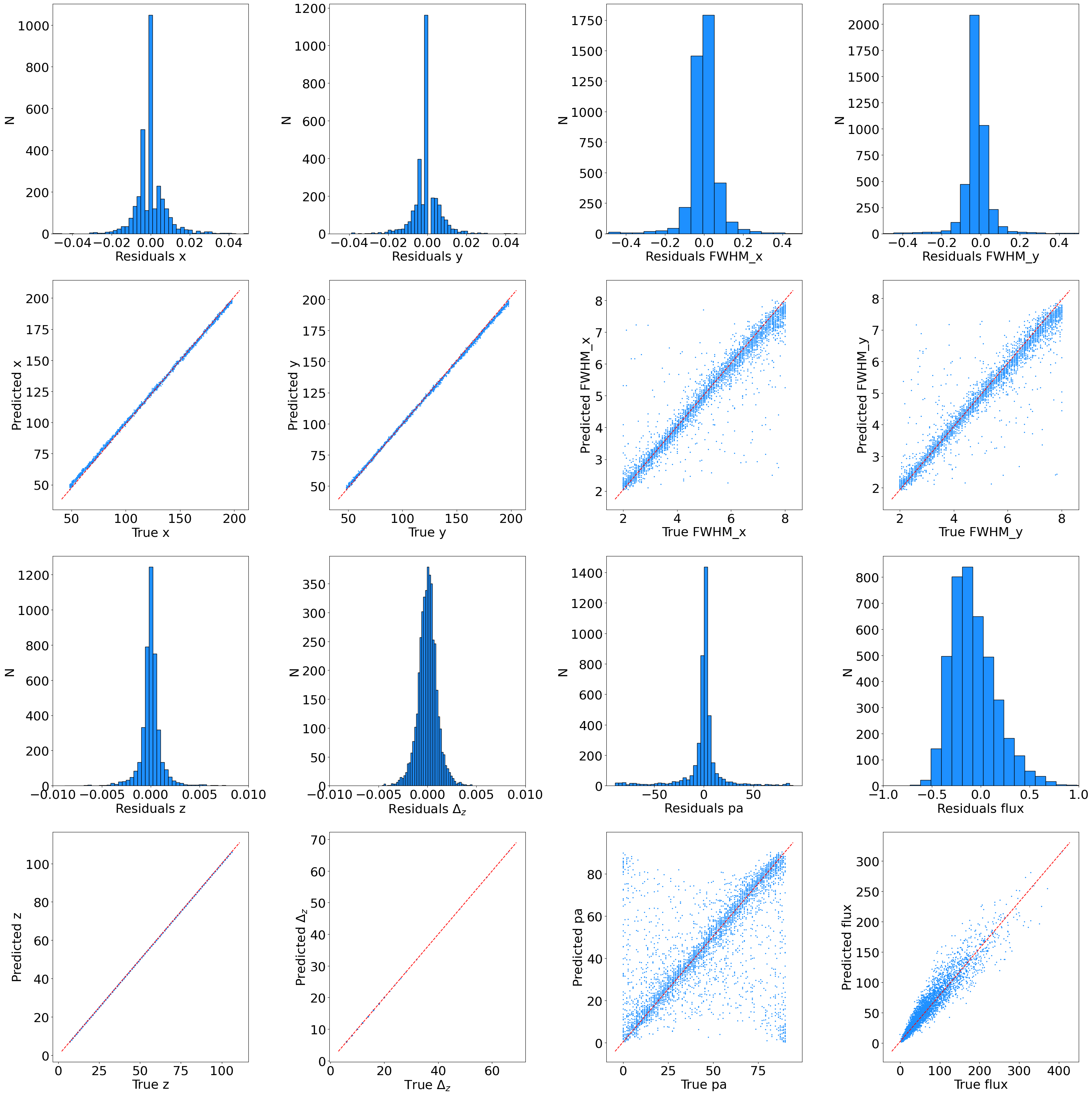}
    \caption{Scatter plots of the true parameters values against the models predictions and the corresponding residuals histograms. The red dotted lines in each scatter plot represent the bisector of the quadrant, i.e. if all instances were perfectly predicted, they would all lie on the red line.}
    \label{fig:morphological}
\end{figure*}

Fig~\ref{fig:morphological} shows the scatter plots of the true parameters versus the predicted ones and the corresponding residuals histograms. The residuals are produced, for each parameter through the following equation:
\begin{equation}
    res_i = (t_i - p_i) 
\end{equation}
where $res_i$ is the residual for the i-th parameter, $t_i$ is the true parameter value, while $p_i$ is the predicted one. Tab.~\ref{tab:morphological} summarises the performances showing, for each parameter, the mean and standard deviation of the residuals distribution.
\begin{table}
    \centering
    \begin{tabular}{l|c|c}
    \hline
    \textbf{Parameter Residual} & \textbf{mean} & \textbf{std} \\
    \hline
    \textbf{x} (pixels)& $-0.004$ & $0.73$ \\
    \textbf{y} (pixels)& $-0.005$ & $0.67$ \\
    \textbf{FWHMx} (pixels) & $-0.04$ & $0.46$ \\
    \textbf{FWHMy}(pixels) & $-0.12$& $0.45$\\
    \textbf{z}(slices) & $0.0$ & $0.003$ \\
    \textbf{$\Delta_z$} (slices) & $0.0$ & $0.001$ \\
    \textbf{pa} (degrees) & $-0.65$& $20.28$ \\
    \textbf{flux} (mJy/beam) & $-9.56$ & $20.08$\\
    \hline
    \end{tabular}
    \caption{This table shows the mean and standard deviation of all the residual distributions between the true target parameters and the predictions made by our pipeline. The $x$ and $y$ positions are computed from Blobs Finder predicted blobs, $z$ positions and extensions $\Delta_z$ are computed from Deep GRU predictions, and the remaining parameters are predicted from the four ResNets. Alongside each parameter, we also indicate their unit of measurement.}
    \label{tab:morphological}
\end{table}

\begin{figure}
	\centering
	\includegraphics[width=\columnwidth]{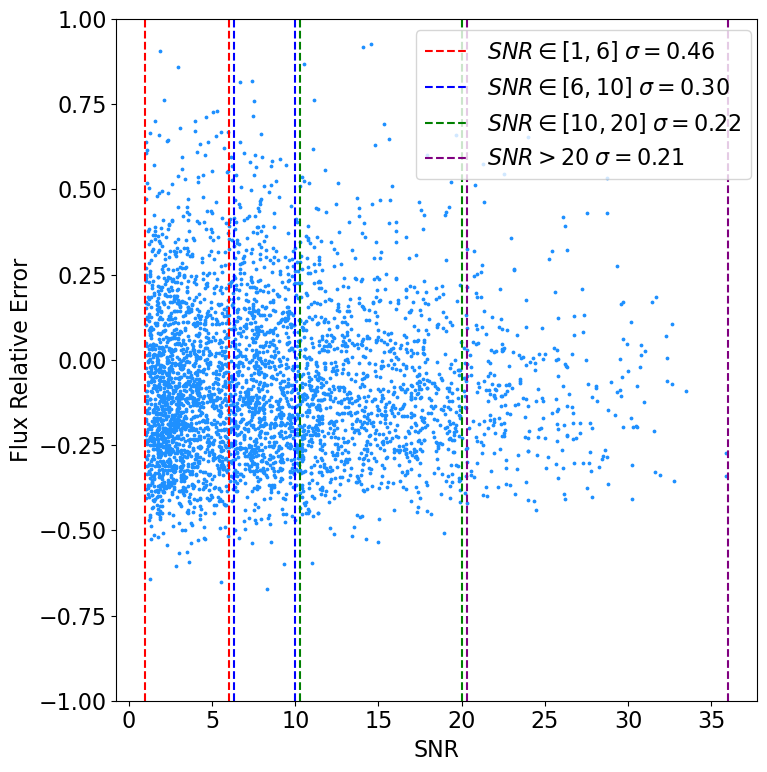}
	\caption{Scatter plot of the sources SNR against their flux densities relative errors. 
	Vertical bars divide the plot in section of SNR. The legend shows, for each SNR interval, the standard deviation of the relative errors.}
	\label{fig:flux_residuals}
\end{figure}
\begin{figure}
	\centering
	\includegraphics[width=\columnwidth]{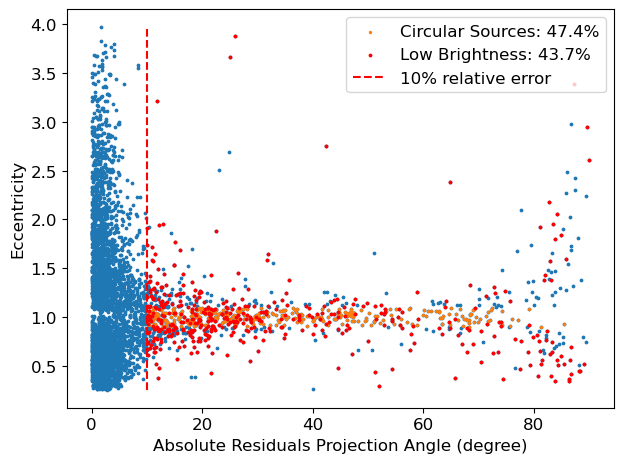}
	\caption{Scatter plot of the sources absolute projection angle residual errors against their eccentricity, defined as the ratio of their FWHMs. The vertical bar delimits the $10\%$ mark
	for the residual error, while the sources highlighted in orange are circular ($e \simeq 1$) and the ones in red have surface brightness lower then $30$ mJy / beam. These sources account for respectively $47.4$ and $43.7$ of all sources with a relative error higher then $10\%$.}
	\label{fig:pa_characterization}
\end{figure}
The source positions (spatial, frequency) and extensions ($FWHM_x$, $FWHM_y$ and $\Delta_z$) are detected with subpixel accuracies.
%As it can be seen, the source positions both spatial and in frequency are detected with subpixel accuracies and the %same can be said for the extensions ($FWHM_x$, $FWHM_y$ and $\Delta_z$). 
 Concerning the performances on the flux densities and projection angle regression, the relative error is defined as follows:
\begin{equation}
    rel_i = (t_i - p_i) / t_i
\end{equation}
where $rel_i$, $t_i$ and $p_i$ are, respectively, the relative error, the true parameter value and the predicted one for the i-th parameter.
The achieved relative errors for the flux densities are $0.07$ with a standard deviation of $0.36$. 
In particular, $68\%$ of sources have a relative error on the flux density which is less than $1\%$ away from the true value, $80\%$ less than $10\%$ and $87\%$ less then $20\%$. Regarding the projection angles, $53\%$ of sources have a relative error which is less then $1\%$ away from the true value, $73\%$ less then $10\%$ and $81\%$ less then $20\%$. The scatter plot of true pa versus predicted ones (Fig.~\ref{fig:morphological} last row, third panel from the left) shows two regions where the nets fail to make accurate predictions, namely, around the values of $0$ and $90$ degrees. This could be expected given the well known degeneracy encountered in measuring projection angles for almost circular sources. 
Fig.~\ref{fig:pa_characterization} shows the scatter plot of the absolute values of the residuals on the projection angle estimation versus the source eccentricity (defined as $e = FWHM_x / FWHM_y$). $47.4\%$ of all sources with a residual error higher than $10\%$ are almost circular ($e \sim 1$) and $43.7\%$ have a surface brightness less than $30$ mJy / beam.
Regarding the surface brightness estimation, the scatter plot of the true flux density versus the predicted one shows that at higher fluxes, the predictive error increases. This behaviour can be explained due to the fact that brightest sources represent only $16\%$ of the data set and that the ResNet is trained, by minimizing the $l_1$ loss, to predict the median of brightness distribution which is $51.2$ in our Train set. The combination of the choice of the loss function and scarcity of bright sources in the data, encourages the model to focus more on less bright sources while treating bright sources as outliers.
Fig.~\ref{fig:flux_residuals} shows the scatter plot of the sources SNR against relative errors on the flux estimations. The standard deviation of 
the flux relative errors distribution halves for sources with SNR higher than $10$ with respect to fainter sources.

\section{Discussion and Conclusions}\label{sec:conclusions}
The transition of many existing and planned radio interferometers to the Terabyte data regime requires the community to develop automatized source detection and characterization pipelines capable to cope with data streams of ever increasing size and complexity. In recent years several attempts have been made to use deep-learning in order to speed up procedures and to make the detection pipeline less subjective. 
In this work we present a novel deep-learning-based source detection and characterization pipeline for radiointerferometric datacubes. This pipeline works directly on dirty interferometric calibrated data cubes which have not undergone any prior de-convolution (i.e., on the results of the Fourier transform applied to the calibrated visibility data) and combines spatial and frequency information to detect and characterize sources within the cubes. 
The pipeline was fine tuned on the characteristics of ALMA data (commonly processed with the \textsc{CASA} software). Nonetheless, given the similarities existing among radio interferometric data from different instruments, the pipeline can be  exported (in presence of a large enough number of simulations for training) to other instruments (such as LOFAR, SKA, VLBI, VLTI).
The proposed pipeline has the potential to support \textsc{CASA} with a new design for  the image reconstruction and/or to provide a speed up procedure for convergence purposes.
Summarising: the pipeline is composed by six deep learning models interconnected through logical operations: Blobs Finder detects sources within the frequency integrated data cubes, 
Deep GRU exploits the frequency domain and detects emission peaks in the spectra extracted from sources detected by Blobs Finder, and the ResNets regress the source parameters from 
'spectrally focused' images created by cropping spatially around the sources, and integrating within their emission range found by Deep GRU, and the line emission images created by masking the cube with the 3D emission models found by combining Blobs Finder and Deep GRU predicted emission ranges (in the spatial and frequency planes of the cube, respectively). 
In order to test the performances of the pipeline, an ALMA observation simulation code was developed. The simulation code is made available through 
GitHub to allow the community to generate thousands of ALMA data cubes in parallel. 
The source detection capabilities of the pipeline can be summarised by the performances achieved on the test set: $92.3\%$ of the simulated sources are detected with no false positives. While the achieved performances are promising for the prospect of applying the pipeline to real data, the low FPRs of the DL models and the subsequent removal of the remaining FPs through SNR and geometrical criteria (FP detection and source deblending step) are closely related to the simplified assumptions that were made to generate the mock data. Integration times and antenna configurations are kept constant resulting in a very low dirty beam variation across the data. Only single peak spectra are simulated without taking into consideration the effects of inclination with the observer line of sight or velocity dispersion within the galaxy, resulting in simplified spectral profiles.
Quality assessment is performed by comparing our results with three other methods, \textsc{blobcat}, \textsc{Sofia-2} and \textsc{decoras}. We notice a substantial improvement in both precision and recall with respect to the first two methods and a smaller improvement with respect to \textsc{decoras}. 
Regarding the source characterization performances, source positions are found with subpixel errors in spatial and frequency domains, while projection angles and flux densities estimations show a relative error within the standard amplitude calibration error of interferometric data ($\simeq 10\%$ \cite{cortes}) for respectively $73\%$ and $80\%$ of all sources.
Regarding the projection angle regression, the ResNet is not able to correctly predict this parameter for circular sources only, being trivial. Given the insights obtained from the analysis of the ResNet performances, a way to avoid the futile regression of the projection angle on circular sources, could be to first compute the source eccentricity (from the predicted $FWHM_x$ and $FWHM_y$), and then regress the projection angle only for non circular sources.
The flux relative error increases with source brightness and decreases with SNR. While explaining the latter trend is easy, the first appears to be counter-intuitive. In fact, the brighter a source is, the easier it should be to measure its flux density. This trend comes from the fact that our simulations contain many more faint sources than bright ones (see Fig.~\ref{fig:simulated_parameters_1}) and from our choice of training the ResNet with an $l_1$ loss.
Regarding the execution time, our pipeline can process a simulated data cube ($67.3$ MBs) in
$\sim 10$ ms. Although the DL pipeline has not yet been tailored for the use of \textsc{CASA}, the technique is capable to speed up \textsc{TCLEAN} by leveraging correlations in all dimensions of the cube.
In future work, we plan to further investigate ways to improve our detection and characterization results. In fact we aim at improving our simulation code to include more complex galaxy morphologies, use physically-based models for the galaxy kinematics, and employ spectral catalogues to generate several spectral profiles for different class of sources with the primary goal of improving the quality of our simulations, and the additional goal of having a publicly available and easy to use simulation code that the community may use to generate common data sets on which compare different architectures.
Alongside we want to modify our pipeline to account for such more complex spectral profiles. The DL pipeline is currently tailored to detect single peak emission lines, assuming that celestial sources have a single positively defined emission peak in their spectra. 
The above does not hold for absorption lines detection or for sources with multi-peak spectral profiles. Both the peak detection (on Deep GRUs denoised spectra) and FP detection and removal (on spectrally focused images through SNR criteria) algorithms will thus need be modified when dealing with real observations.
We are also planning to make an assessment on faint signal detection especially in the presence of strong sidelobes in the cubes and asses the pipeline capabilities in the case of data containing several uv coverages and array configurations, which should result in a greater variation of the dirty beam within the data, thus posing a more complex image reconstruction problem. Furthermore, we also plan to make tests about incorporating the dirty beam within our pipeline in order to improve the image reconstruction capabilities, and to test the FP removal pipeline against a DL classifier based on our ResNet architecture in case of the aforementioned more complex data.
The DL pipeline is going to be further extended on continuum imaging and applied to real ALMA observations with the aim of finding new faint serendipitous galaxies in the neighbourhood of brighter companions. This task has been proven difficult for classical algorithms.

%%%%%%%%%%%%%%%%%%%%%%%%%%%%%%%%%%%%%%%%%%%%%%%%%%
\section*{Data and Software Availability}
The simulation code to generate the data used in this work is made publicly available through Github (ALMASim \href{https://github.com/MicheleDelliVeneri/ALMASim}{\faGithub}). Detailed instruction on how to setup a python environment with all the pip packages needed to run the code and generate the data are outlined in the repository. The simulation code has been tested only on Unix distributions (Ubuntu, CentOS, macOS) and is composed by two bash scripts and several python scripts which generate the sky models, run the CASA tasks to simulate ALMA observations and produce the cubes, control the noise properties, and record the source parameters. The bash scripts are written for the IBISCO architecture which uses the slurm workload manager to split computations among nodes, and thus they should be changed to reflect the architecture on which they are executed. On our system, $5,000$ cube pairs were generated on $400$ Intel Xeon E5-2680 CPUs in around a day. The pipeline is also implemented in python and made publicly available through GitHub (DeepFocus \href{https://github.com/MicheleDelliVeneri/DeepFocus}{\faGithub}). The Deep Learning models, the data loading, augmentation, training and testing routines are written through the pytorch library. While the simulation code is fully documented, at the time of writing of this paper, the documentation for the pipeline is still under construction. Basic instructions on how to run the training and testing of the pipeline are outlined on the GitHub page. Also in this case, the script parameters are tailored for the IBISCO architecture and should be changed accordingly. Blobs Finder' training lasted $\sim 4$ hours on a $2$ NVIDIA Tesla K20 and it made the predictions on the test set in $\sim 23$ seconds including I/O operations. While Blobs Finder was trained on two GPUS, Deep GRU and the ResNets were trained on a single NVIDIA Tesla K20. Regarding the name of the pipeline, given its modularity, it will probably not hold for long and will be reformulated once more extensive tests are made over the several problems which are investigated in this paper: image reconstruction, source detection and source characterization.
Deep GRU's training lasted $\sim 20$ minutes on spectra produced from the truth catalogue and $\sim 3$ minutes on Blobs Finder' predictions on the training set, and it made the predictions on the test set in $\sim 9$ seconds including I/O operations. Each ResNet's training lasted $\sim 2$ hours on spectrally focused images (or line emission images for flux density regression) produced from the truth catalogue, $\sim 30$ minutes on spectrally focused images produced from Blobs Finder and Deep GRU predictions, and predictions on the test set were made in $\sim 5$ seconds including I/O operations.

\section*{Acknowledgements}
The authors wish to thank the referee for his very valuable suggestions who greatly improved the readability of the paper and made it more robust. We want to thank The IBISCO HPC admin group for allowing us to extensively use their cluster to simulate the data and carry out all our experiments. MDV wishes to thank the European Southern Observatory (ESO) and the Department of electric Engineering and Information Technologies at the University Federico II for partial financial support. This work is supported by an ESO internal ALMA development study investigating interferometric image reconstruction methods and was partially sponsored by EU through the SUNDIAL ITN.   
%--------------------------------------------------------- NOTES --------------------------------------------------------------



\bibliographystyle{mnras}
\bibliography{main.bib} % if your bibtex file is called example.bib

%%%%%%%%%%%%%%%%%%%%%%%%%%%%%%%%%%%%%%%%%%%%%%%%%%

%%%%%%%%%%%%%%%%% APPENDICES %%%%%%%%%%%%%%%%%%%%%

%\appendix



%%%%%%%%%%%%%%%%%%%%%%%%%%%%%%%%%%%%%%%%%%%%%%%%%%

% Don't change these lines
\bsp	% typesetting comment
\label{lastpage}
\end{document}